\documentclass[a4paper,usenatbib,useAMS]{mnras}
\usepackage{graphicx}
\usepackage{lscape}
\usepackage{perpage}
\usepackage{amsmath}
\usepackage{amssymb}
\usepackage{color}
\MakePerPage{footnote}

\newcommand{\beq}	{\begin{equation}}
\newcommand{\eeq}	{\end{equation}}
\newcommand{\beqa}	{\begin{eqnarray}}
\newcommand{\eeqa}	{\end{eqnarray}}
\newcommand{\e}	        {$^{-1}$}
\newcommand{\ee}	{$^{-2}$}

\newcommand{\calm}	{{\cal M}}

\newcommand{\msun}      {\rm M_\odot}

\newcommand{\vecB}	{{\bf B}}

\newcommand{\vecv}	{{\bf v}}
\newcommand{\vecx}	{{\bf x}}

\newcommand\fnt		{\footnotetext}

\newcommand{\alfven}    {{Alfv$\acute{\rm e}$n }}

\newcommand{\alfvenic}  {{Alfv$\acute{\rm e}$nic }}
\newcommand{\alfvenicstop}  {{Alfv$\acute{\rm e}$nic}}

\newcommand{\muphi}	{\mu_{\Phi}}

\newcommand{\avir}      {\alpha_{\rm vir}}
\newcommand{\avirf}      {\alpha_{\rm vir,\,f}}

\newcommand\bpos        {B_{\rm POS}}

\newcommand{\ma}	{{\calm_{\rm A}}}

\newcommand{\mug}	{$\mu$G}
\newcommand{\muh}   {\mu_{\rm H}}
\newcommand{\mmug}	{\mu{\rm G}}

\newcommand\cs		{c_{\rm s}}

\newcommand\POS         {{\rm POS}}

\newcommand{\bosf}       {B_{0,\rm SF}}
\newcommand{\crit}		{{\rm crit}}

\newcommand{\muphipos}		{\mu_{\Phi,\POS}}

\newcommand{\svt}       {\sigma_{V,\,\rm tot}}

\newcommand{\vecl}      {\boldsymbol{\ell}}

\title[Role of Magnetic Fields in Filamentary Clouds]{The Role of Magnetic Fields in the Stability and Fragmentation of Filamentary Molecular Clouds: Two Case Studies at OMC-3 and OMC-4}
\author[Pak Shing Li et al.]{
Pak Shing Li$^{1}$\thanks{E-mail:psli@berkeley.edu} (PSL),
Enrique Lopez-Rodriguez$^{2}$, 
Archana Soam$^{3}$,
Richard I. Klein$^{1,4}$
\\
$^{1}$Astronomy Department, University of California, Berkeley, CA 94720\\
$^{2}$Kavli Institute for Particle Astrophysics and Cosmology (KIPAC), Stanford University, Stanford, CA 94305, USA \\
$^{3}$ Indian Institute of Astrophysics, II Block, Koramangala, Bengaluru 560034, India\\
$^{4}$Lawrence Livermore National Laboratory,P.O.Box 808, L-23, Livermore, CA 94550\\
}

\begin{document}

\date{}

\pubyear{}

\label{firstpage}
\pagerange{\pageref{firstpage}--\pageref{lastpage}}
\maketitle

\begin{abstract}
We present the stability analysis of two regions, OMC-3 and OMC-4, in the massive and long molecular cloud complex of Orion A. We obtained $214~\mu$m HAWC+/SOFIA polarization data, and we make use of archival data for the column density and C$^{18}$O (1-0) emission line. We find clear depolarization in both observed regions and that the polarization fraction is anti-correlated with the column density and the polarization-angle dispersion function. We find that the filamentary cloud and dense clumps in OMC-3 are magnetically supercritical and strongly subvirial. This region should be in the gravitational collapse phase and is consistent with many young stellar objects (YSOs) forming in the region. Our histogram of relative orientations (HROs) analysis shows that the magnetic field is dynamically sub-dominant in the dense gas structures of OMC-3. We present the first polarization map of OMC-4. We find that the observed region is generally magnetically subcritical except for an elongated dense core, which could be a result of projection effect of a filamentary structure aligned close to the line-of-sight. The relative large velocity dispersion and the unusual positive shape parameters at high column densities in the HROs analysis suggest that our viewing angle may be close to axes of filamentary substructures in OMC-4. The dominating strong magnetic field in OMC-4 is unfavorable for star formation and is consistent with much fewer YSOs than in OMC-3.

\end{abstract}


\section{Introduction}
How important magnetic fields are in star formation is one of the main questions toward a complete picture of star formation theory. With the current increase in polarization mapping of molecular clouds, where stars form, available in the last decade, we can also address the question on how dynamically important the magnetic fields are in the formation of molecular clouds. The current consent is that magnetic fields are dynamically significant in molecular clouds. This result is based on a) the comparison of the mean field orientation inside molecular clouds to that of large-scale field in the diffuse interstellar medium (ISM) \citep[e.g.][]{li09,li11}, and b) the histogram of relative orientations (HRO) analysis \citep[e.g.][]{sol13,sol17,fis19} that reveals the change of alignment of magnetic fields with respect to the filamentary structures as column density increases.

Polarization observations of magnetic fields around filamentary molecular clouds show that magnetic fields are mostly perpendicular to the long axis of large, $\ge$ few pc, filamentary molecular clouds \citep[e.g.][]{cha11,and14,pla16,liu18,sug19}. There are also evidences that magnetic fields are mostly perpendicular to the dense filamentary substructures inside molecular clouds \citep[e.g.][]{sol13,sol17,fis19,sei20}, while low column density filaments are mostly observed parallel to the magnetic field lines \citep[e.g.][]{pal13,pla16}. \citet{and14} proposed that this is a result of gas accretion through the modulation of magnetic field onto dense structures, as seen in numerical simulations with moderately strong magnetic field \citep[e.g.][]{bas09,li10,li15,ino18}. The three mechanisms for filament formation identified by \citet{abe21}, including the fast shock wave, compressive flows, and gravity, effectively function in the presence of moderately strong magnetic field when gas flows easier along the field and results in magnetic field perpendicular to the filament axis.

Recent observational results from BLASTPol \citep[e.g.][]{fis19} show that high resolution small-scale magnetic field structures inside the molecular cloud Vela C is more complex than the large-scale external magnetic field. The orientations of the small-scale magnetic fields depend on the orientations of dense substructures inside the cloud. Gravity of dense structures and gravity-driven turbulence become important in competing with magnetic field and strongly perturb the magnetic field, which is shown in numerical simulations \citep[e.g.][]{li19} and has been seen in recent observations \citep[e.g.][]{pil20,arz21}.

\begin{table*}
\caption{Summary of OTFMAP polarimetric observations. From left to right: Object name, date of observations, flight ID, altitude during observations, speed of scan, amplitude of individual scans, duration of exposure of individual scans, number of sets per night, total on-source time per night.}
\label{tab:ObsLOG}
\begin{tabular}{p{2.2cm}cccccccccc}
\hline
\hline
Object& Date	&	Flight	&	Altitude	&
Scan  &  Scan   & Scan  & \#Sets & t  \\
& 	&		&		&
 Rate &   Amplitude  &  Duration &  &  &  &  \\
  & (YYYYMMDD)	&		&	(kft)	&
(\arcsec/sec) &  (EL $\times$ XEL; \arcsec) & (s)  & & (s)\\
\hline
OMC-3 (G208.68-19.20)   &      20200923    &   F689    &   43-44 & 200  & 100$\times$100  &   120 &   6   &   2880    \\
                &    20200924    &   F690    &   43-44 & 200  &
100$\times$100  &   120 &   3   &   1440    \\
OMC-4 (G209.29-19.65)   &      20200923    &   F689    &   43-44 & 200 &   150$\times$150  &   90 &   3   &   1080 \\
                &     20200924    &   F690    &   43-44 & 200  &
150$\times$150  &   120 &   5   &   2400    \\
\hline
\hline
\end{tabular}
\end{table*}

Long filamentary structures, especially those more than several parsec long, are expected to be unstable in the highly turbulent interstellar medium \citep[e.g.][]{li19}. However, these filamentary clouds are commonly seen in giant molecular clouds (GMCs) using {\it Herschel} and {\it Planck} observations, and they appear not to be short-term transient objects. \citet{li19} show that with a moderately strong large-scale magnetic field with \alfven Mach number $\ma \approx 1$, long and slender filamentary clouds can exist for at least 0.9 Myr. With strong turbulence flows from different directions, long filamentary gas will be easily sheared or buckled. The key reason for the longevity of the filamentary cloud in the simulation is the presence of the moderately strong large-scale magnetic field. The strong field reinforces filamentary gas from quick destruction and helps the continuous feeding of gas to the cloud along the field. The simulation shows that magnetic fields play an important role in the formation and maintenance of long filamentary clouds. In the simulation, fragmentation happens in the small filamentary substructures of less than a parsec long and the width $\lesssim 0.1$ pc inside the molecular clouds, where gravity and turbulence are playing equally important, if not more important, roles at the scales of dense clumps and cores.

Orion A is a well studied part of a large molecular cloud complex. Large rings of dust are revealed in the mapping of Orion cloud complex by \citet{sch15}. Orion A may be a part of an ancient bubble of about 100 pc in diameter in the interstellar medium. Visually, Orion A is one of the long molecular clouds near the edge of the bubble and composed of many filament structures. Recent filamentary structure identification is attempted by \citet{zhe21}. They identified 225 fiber-like filaments from an H$_2$ map constructed using $^{12}$CO, $^{13}$CO, and C$^{18}$O observations. The northern part of the Orion A cloud, including regions OMC-1 to OMC-4, appears as an integral shape. The H$_2$ column density map recently obtained by \citet{kon18} showed that the column density is not uniform along the cloud and there are many well defined clumps at different locations along the cloud. The local velocity dispersions they obtained from the $^{12}$CO(1-0), $^{13}$CO(1-0), and C$^{18}$O(1-0) are mostly between 0.5 - 1.5 km s$^{-1}$ and can be as high as 2 km s$^{-1}$. Using N$_2$H$^+$ (1-0) emission map from ALMA mosaics with previous IRAM 30m observations, \citet{hac18} identified 55 dense fiber-like structures in the northern integral-shaped-filamentary part of Orion A. About 50 per cent of the fibers could be gravitationally unstable and have the ratio of mass per unit length to the critical value of an infinite filament in hydrostatic equilibrium between $0.5 \sim 1.5$. The other half are stable with the ratio $< 0.5$.
However, \citet{hac18} and \citet{zhe21} do not have information on magnetic field of the substructures they identified. Support by a perpendicular magnetic field and thermal/turbulent motions can allow a larger line mass for a gravitational stable filamentary cloud \citep[e.g.][]{kas21,li22}.

Orion A with rich filamentary structures allows us to investigate the role of magnetic field in filamentary structure formation and fragmentation. Polarization observations have been carried out at different locations in Orion A before \citep[e.g.][]{mat01,hou04,mat05}. \citet{pat17} obtained the polarization map of the OMC-1 and estimated the magnetic field strength on the plane-of-sky (POS) $\bpos = 1.9 - 11.3$ mG using the classical Davis-Chandrasekhar-Fermi (DCF) method \citep{dav51,cha53,gue21}. The large range of their results is coming from a conservative estimation of uncertainty and it is consistent with estimation from other measurement techniques \citep{coh06,hil09,tan10} at the order of a few milligauss. Using a variant of the DCF method based on the structure function of polarization map (DCF/SF method), \citet{hou09} estimated an B-field strength of $\bpos = 0.76$~mG for OMC-1. \citet{mat01} did not estimate the field strength of the OMC-3 region but \citet{mat05} used the DCF method and estimated the magnetic field strength of the MMS 6 core of OMC-3 to be $\bpos = 0.64$~mG. There is no polarization mapping for OMC-4 region yet.

In this paper, we present our polarization observations of OMC-3 and OMC-4 using the High-resolution Airborne Wideband Camera-plus \citep[HAWC+;][]{vai07,dow10,har18} on-board the 2.7-m Stratosphere Observatory for Infrared Astronomy (SOFIA). With the optical depth mapping of Orion A by \citet{lom14} and the recent publicly released CO observation data by \citet{kon21}, we investigate the physical environment in OMC-3 and OMC-4 regions and determine how magnetic field is shaping the formation of the filamentary clouds and dense clumps in these two regions. In Section 2, we present the HAWC+ polarization mapping method and results, as well as the archival data for column density and velocity dispersion. In Section 3, we use DCF and DCF/SF methods to estimate the magnetic field strengths of the two regions and determine their physical conditions. We also compare our estimation with previous field strength estimations. We present our HRO analysis results in Section 4 and discuss our study on the depolarization in the two observed regions in Section 5. Section 6 is our conclusions on this work. 

\begin{figure*}
\includegraphics[scale=0.95]{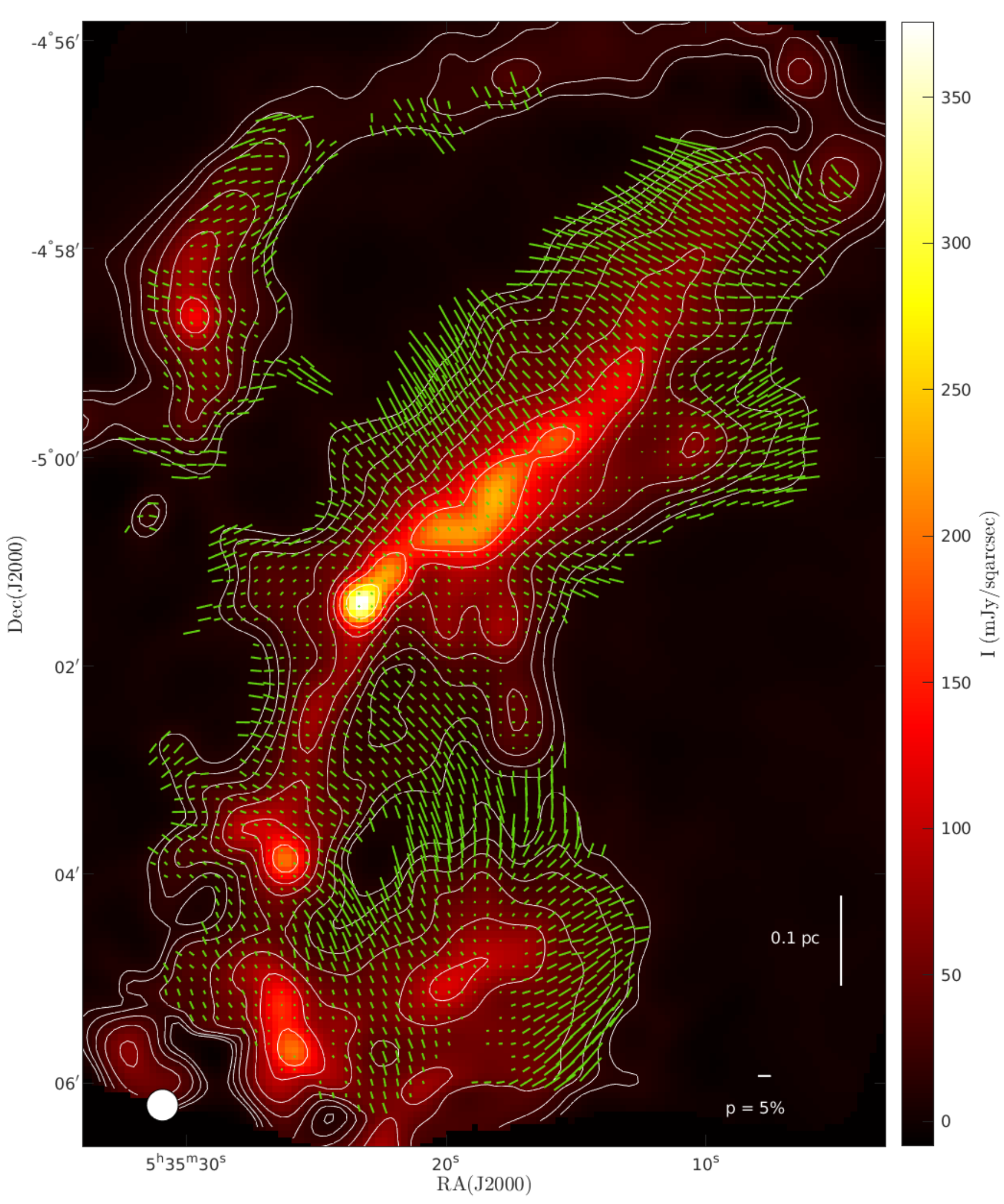}
\caption{Magnetic field orientation map of OMC-3. Total surface brightness (colorscale) at $214$ $\mu$m within $7\arcmin\times10\arcmin$ region using the OTFMAP observations. Contours are plotted at $\log I$~(mJy/sqarcsec) = 1 to 2.4 with a step of 0.2. The starting contour is about $109\sigma$, with $\sigma = 0.092$ mJy/sqarcsec. Polarization measurements at the Nyquist sampling (green lines) have been rotated by $90^{\circ}$ to show the inferred magnetic field orientation. The length of polarization measurements are proportional to the degree of polarization. Only polarization measurements with $P/\sigma_{P}\ge3$, $P\le30\%$, and $I/\sigma_{I} \ge 100$ are shown. The legends of $5$\% polarization and spatial scale of 0.1 pc are shown at the bottom right. The beam size of $18\farcs2$ is shown at the bottom left. 
\label{fig:OMC3_fields}}
\end{figure*}

\begin{figure*}
\includegraphics[scale=0.95]{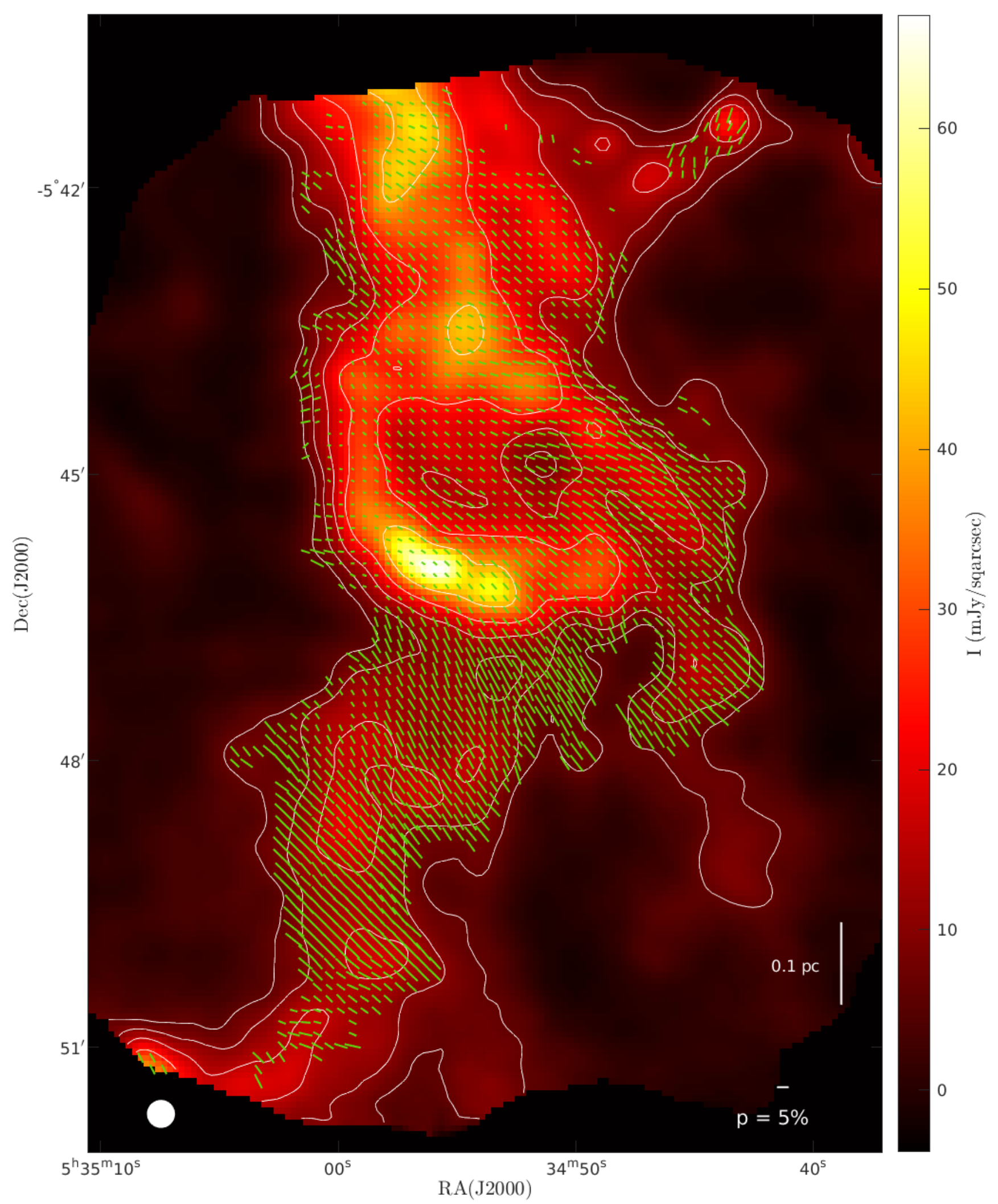}
\caption{Magnetic field orientation map of OMC-4. Total surface brightness (colorscale) at $214$ $\mu$m within $7\arcmin\times10\arcmin$ region using the OTFMAP observations. Contours are plotted at $\log I$ (mJy/sqarcsec) = 0.8 to 1.8 with a step of 0.2. The starting contour is about $73\sigma$, with $\sigma = 0.086$ mJy/sqarcsec.  Polarization measurements at the Nyquist sampling (green lines) have been rotated by $90^{\circ}$ to show the inferred magnetic field orientation. The length of polarization measurements are proportional to the degree of polarization. The legends of $5$\% polarization and spatial scale of 0.1 pc is shown at the bottom right. Only polarization measurements with $P/\sigma_{P}\ge3$, $P\le30\%$, and $I/\sigma_{I} \ge 100$ are shown. The beam size of $18\farcs2$ is shown at the bottom left. 
\label{fig:OMC4_fields}}
\end{figure*}

\section{Observations and data reduction}

\subsection{Polarization mapping using HAWC+/SOFIA}

OMC-3 (G208.68-19.20) and OMC-4 (G209.29-19.65) were observed  (ID: 08\_0027; PI: Li, P.S.) using HAWC+/SOFIA. HAWC+ polarimetric observations simultaneously measure two orthogonal components of linear polarization arranged in two arrays of $32 \times 40$ pixels each. Both objects were observed at $214$ $\mu$m, where HAWC+ has a detector pixel scale of $9$\farcs$37$ pixel$^{-1}$, and beam size (full width at half maximum, FWHM) of $18$\farcs$2$. For a distance to Orion A $\sim 400$ pc \citep[e.g.][]{men07,sch15,kou17}, the beam size corresponds to $\sim 0.035$ pc. We shall use 400 pc for the distance to OMC-3 and OMC-4 in the analysis.

All observations were performed using the on-the-fly-map (OTFMAP) polarimetric mode and high-level data products were delivered by the SOFIA Science Center\footnote{HAWC+ data reduction details can be found at: \url{https://www.sofia.usra.edu/sites/default/files/2022-02/hawc_users_revJ_0.pdf}}. The OTFMAP polarimetric mode has been successfully applied to L1495/B211 \citep{li22}. We point out the reader to \citet{lop22} for the full characterization of the OTFMAP polarimetric mode of HAWC+. Data were reduced by the SOFIA Science Center using the \textsc{hawc\_drp\_v2.5.0} pipeline. In summary, the OTFMAP polarimetric observations are performed using a sequence of four Lissajous scans, where each scan has a different halfwave plate (HWP) position angle (PA) in the following sequence: $5^{\circ}$, $50^{\circ}$, $27.5^{\circ}$, and $72.5^{\circ}$. The telescope is driven to follow a parametric curve with a nonrepeating period whose shape is characterized by the relative phases and frequency of the motion, i.e. Lissajous pattern. Each scan is characterized by the scan amplitude, scan rate, scan angles, and scan duration. A summary of the observations are shown in Table \ref{tab:ObsLOG}. The Stokes IQU parameters using the double difference method in the same manner as the standard chop-nod observations carried by HAWC+ described in Section 3.2 by \citet{har18}. The degree (P) and polarization angle (PA) of polarization were corrected by instrumental polarization (IP) estimated using OTFMAP polarization observations of planets is the polarization uncertainty. To ensure the correction of the PA of polarization of the instrument with respect to the sky, the scans were taken with a fixed line-of-sight (LOS) of the telescope. The Stokes QU were rotated from the instrument to the sky coordinates. The polarization fraction was debiased, i.e. $p' = \sqrt{P^{2}-\sigma_{p}^{2}}$, where $\sigma_{p}$ is the polarization uncertainty, and corrected by polarization efficiency.  Final images of OMC-3 and OMC-4 have a total on-source time of 4320s and 3480s, respectively, and the observing logs are shown in Table \ref{tab:ObsLOG}. The delivered reduced data have a pixel scale of $3$\farcs$70$ pixel$^{-1}$, which corresponds to 1/4 beam size, and smoothed using a 2D Gaussian profile with a FWHM equal to the beam size of the $214$ $\mu$m. The pixels within the beam size are correlated and polarization maps are shown using Nyquist sampling. Our analysis is performed using the half-beam measurements, i.e. Nyquist sampling. Figures \ref{fig:OMC3_fields} and \ref{fig:OMC4_fields} show the final reduced images of OMC-3 and OMC-4, respectively, with their polarization measurements rotated by $90^{\circ}$ to display the B-field orientation. Our observations have a sensitivity of $\sigma_{I} = 0.044$ mJy/sqarcsec and $\sigma_{PI} = 0.056$ mJy/sqarcsec in total and polarized flux density, respectively for OMC-3. OMC-4 observations have a sensitivity of $\sigma_{I} = 0.056$ mJy/sqarcsec and $\sigma_{PI} = 0.068$ mJy/sqarcsec in total and polarized flux density, respectively. Only polarization measurements with $P\le30$\% are used, which upper-limit is given by the maximum polarized emission of $\sim25$\% found by \textit{Planck} observations \citep{pla13}.

\subsection{Archival data}
\label{subsec:ArchivalData}

To support the analysis of the velocity dispersion, we make use of observations of several molecular lines. We use the CARMA-NRO Orion high-resolution survey data from \citet{kon18} with a beam size of $10\arcsec\times8\arcsec$ for $^{12}$CO(1-0) and $8\arcsec\times6\arcsec$ for $^{13}$CO(1-0) and C$^{18}$O(1-0), better than the resolution of the HAWC+ polarization map. The velocity channel widths are 0.25 km s$^{-1}$ for $^{12}$CO(1-0) and 0.22 km s$^{-1}$ for $^{13}$CO(1-0) and C$^{18}$O(1-0). The rms noise per channel is 0.86, 0.64, and 0.47 K for $^{12}$CO(1-0), $^{13}$CO(1-0), and C$^{18}$O(1-0), respectively \citep[see table 2 in][]{kon18}. We convolve and reproject the molecular line data to the same resolution as the HAWC+ data. Specifically, the integrated emission line (moment 0) and velocity dispersion (moment 2) maps were smoothed using a Gaussian profile with a FWHM equal to the resolution of the HAWC+ observations. Finally, the smoothed images were projected to the HAWC+ pixel grid (Figure \ref{fig:correlation}). We only use the same LOS molecular line measurements associated with the polarization measurements shown in Figure \ref{fig:OMC3_fields}.

To support the analysis of the column density, we infer the density from the optical depth measurements by \citet{lom14}. They used the \textit{Planck} and \textit{Herschel} dust-emission data and 2MASS NIR dust-extinction data to obtain the optical depths of the entire Orion complex. The resolution of their map of optical depth is $36\arcsec$. Following \citet{lom14}, we adopt $\Sigma/A_K \simeq 183~\msun\;{\rm pc}^{-2}$, where $\Sigma$ is the surface density and $A_K$ is the K-band extinction, so that the hydrogen column density is
\beq
N({\rm H})=\frac{\Sigma}{\mu_{\rm H}} = 1.634 \times 10^{22} \times A_K~{\rm cm}^{-2},
\eeq
where $\mu_{\rm H} = 2.34\times10^{-24}$~g is the mass per hydrogen nucleus. \citet{lom14} fitted a linear relationship between optical depth $\tau_{850}$ and $A_K$, and obtained $A_K = \gamma \tau_{850} + \delta$, where $\gamma = 2640$ mag and $\delta = 0.012$ mag for Orion A. We construct the column density map based on their $\tau_{850}$ data. For reference, \citet{sch21} published their latest observation of the northern part of Orion A with higher resolution at $8\arcsec$. We find that their column density map matches the map by \citet{lom14} well but \citet{sch21} did not cover the OMC-4 region. For consistency, we have decided to use the results from \citet{lom14}.

\section{Physical States of OMC-3 and OMC-4}

\subsection{Magnetic field estimation using DCF methods}
\label{sec:dcf}

The original DCF method for estimating the magnetic field strengths in the ISM is based on the assumptions that the medium is isotropic and that variations in the orientation of the field are due to \alfven waves. For \alfven waves the equation of motion implies
\beq
\delta\vecv=\pm\frac{\delta\vecB}{(4\pi\rho)^{1/2}},
\label{eq:dv}
\eeq
where $\delta \vecv$ and $\delta\vecB$ represent the wave amplitude in the POS. In the linear regime, it implies equipartition between the turbulent kinetic energy of motions normal to the mean magnetic field in the POS, $\vecB_0$. The corresponding field energy in the waves is $\rho\delta v_\perp^2/2=\delta B_\perp^2/8\pi$, where the POS quantities $\delta \vecv_\perp$ and $\delta\vecB_\perp$ are perpendicular to the mean POS field. Under the assumption that the turbulent velocities are isotropic, the rms value of $\delta v_\perp$ is the same as the LOS velocity dispersion, $\sigma_V$. We use a version of this method that is valid for larger dispersions of PAs than the original method \citep{fal08, li22},
\beqa
B_0&=&f_{\rm DCF} \,\frac{(4\pi\rho)^{1/2}\sigma_V}{\tan\sigma_\theta},\\
&=& 0.0857 \sqrt{n_5({\rm H})}\,
\frac{\sigma_V}{\tan\sigma_{\theta}}  ~~~ {\rm mG},
\label{eq:bpos2}
\eeqa
where $n_5({\rm H})$ is the number density of hydrogen nuclei in units of $10^5$~cm$^{-3}$, $\sigma_V$ is the non-thermal velocity dispersion measured in km~s\e, and $\sigma_{\theta}$ is the dispersion of the PAs. We have set the factor $f_{\rm DCF}$, which corrects for the approximations made in deriving the DCF relation, to be 0.5 based on the results of \citet{ost01}. Comparison with numerical simulations confirms that this formula (with $\tan\sigma_\theta$ replaced by $\sigma_\theta$ in radians under the assumption that $\sigma_\theta$ is small) is valid when $\sigma_\theta \leq 25^\circ$ \citep{ost01}. The latter relation (with $\sigma_\theta$) is often used for larger dispersions, however. Since polarization angles are restricted to lie in the range $-90^\circ< \theta\leq 90^\circ$, their values depend on the orientation of the coordinate system. We choose the coordinate system that minimizes $\sigma_\theta$, as recommended by \citet{pad01}.

The turbulent component of the POS field perpendicular to the mean field is
\beq
B_{t\perp}=B_0\tan\sigma_\theta,
\label{eq:btp}
\eeq
which was denoted $\sigma_{\delta B_\perp}$ by \citet{li22}. 
The errors in the approximations that went into Equation (\ref{eq:btp}) are (a) of order $(B_{t\perp}/ B_0 )^4$ and (b) less than 11 per cent for $\theta < 90^\circ$. It is also based on the assumption that the true field angles are approximately equal to the PAs, which is assumed in all DCF treatments. The net effect of the approximations is corrected by the factor $f_{\rm DCF}$, which is set from simulations. If the turbulent field is isotropic, the total POS turbulent field is $B_t=\surd 2 B_{t\perp}$ and the total 3D turbulent field is $\surd 3 B_{t\perp}$.

We estimate $\sigma_\theta$ as the weighted standard deviation of polarization angles in each case: 
\begin{equation}
{\sigma_\theta^2  } = \frac{N}{N-1}\,\frac{1}{w}\, 
\sum\limits_{i=1}^N w_i \, (\theta_i - \bar{\theta}_w)^2 , 
\label{eq:sigmatheta}
\end{equation}
\noindent
where $N$ is number of pixels in the region, $w_i = 1/\sigma_i^2 $ the weight of measurement $i$ given the measurement error in PA $\sigma_i$, $w = \sum\limits_{i=1}^N w_i$, and $\bar{\theta}_w = (1/w)\, \sum\limits_{i=1}^N w_i\, \theta_i$ is the weighted mean polarization angle in the region. The dispersion of the column density, $N({\rm H})$, and velocity dispersion, $\sigma_V$, are obtained using the same weighting as in Equation (\ref{eq:sigmatheta}) and shown as the uncertainties for the means in Table \ref{tab:properties}.

\begin{figure*}
\includegraphics[angle=0,scale=0.79]{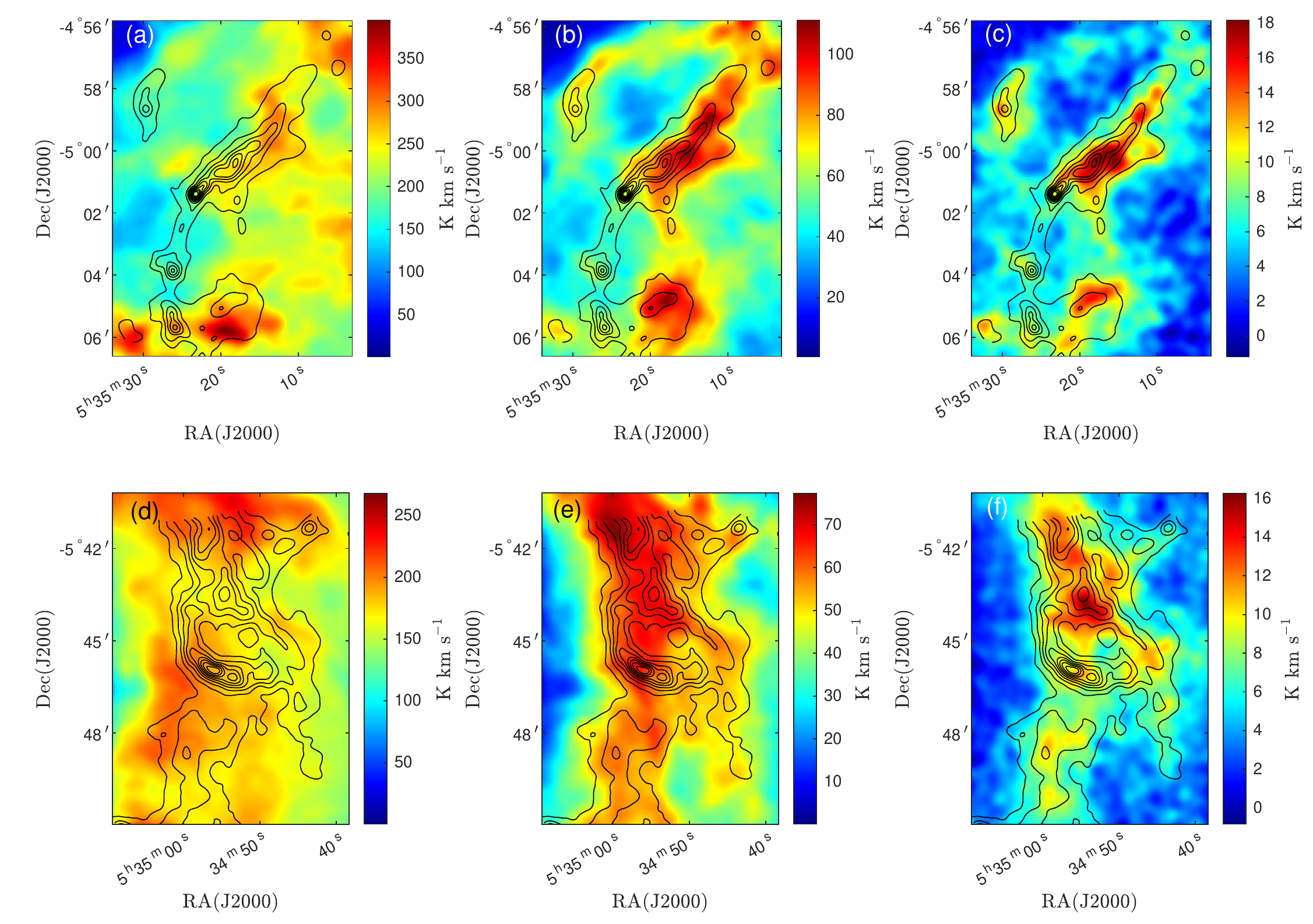}
\caption{Top three panels are the 0-moment maps of molecular lines (a) $^{12}$CO(1-0), (b) $^{13}$CO(1-0), and (c) C$^{18}$O(1-0) of OMC-3 region from \citet{kon18} with contours of Stokes I from HAWC+ polarization observation. The same for the bottom three panels are the 0-moment maps of molecular lines (d) $^{12}$CO(1-0), (e) $^{13}$CO(1-0), and (f) C$^{18}$O(1-0) of OMC-4 region.
\label{fig:correlation}}
\end{figure*}

\begin{figure*}
\includegraphics[scale=0.7]{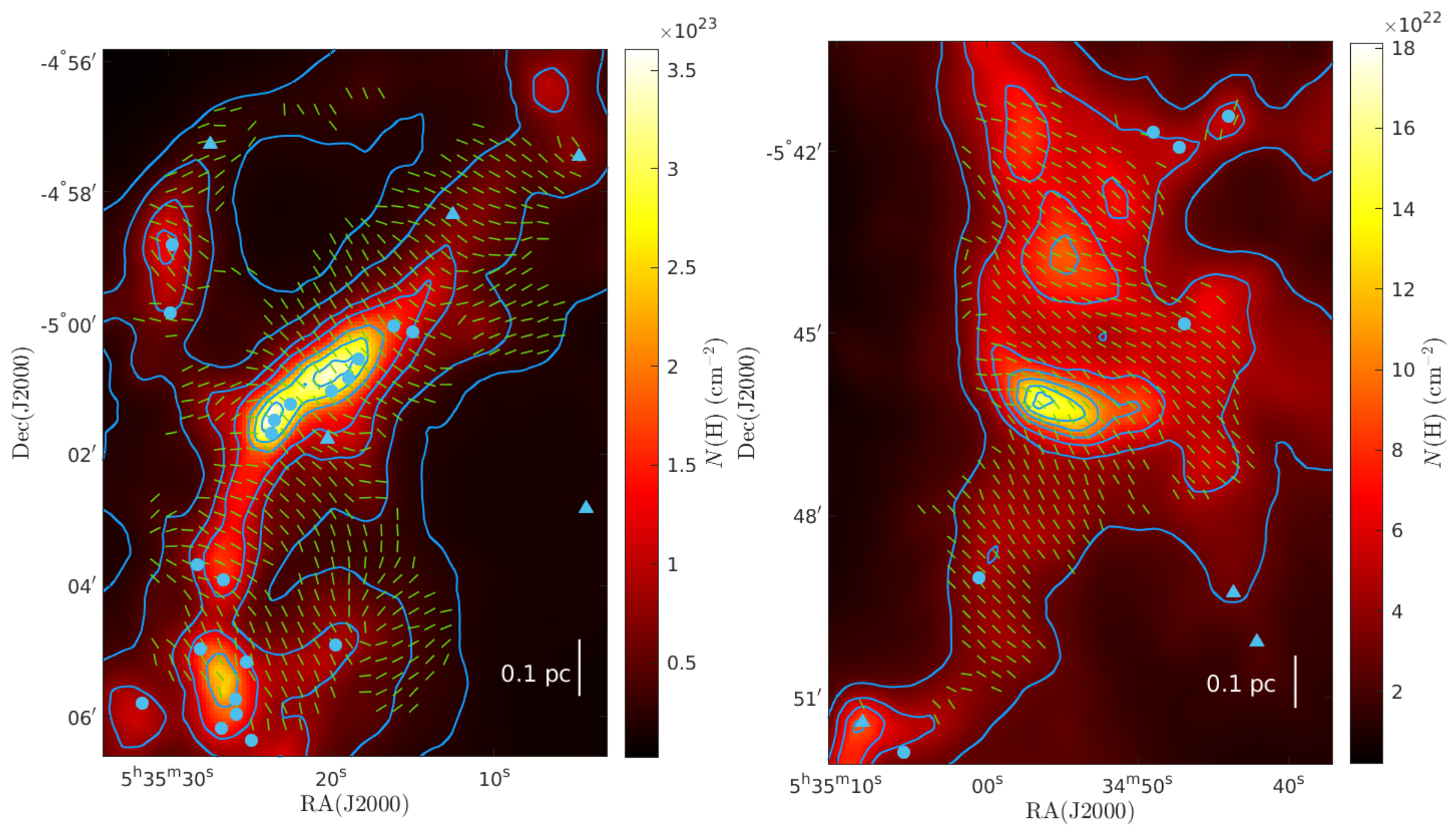}
\caption{The polarization map of OMC-3 (left) and OMC-4 (right) on top of the column density (colour scale) map from \citet{lom14}. The inferred magnetic field orientations are shown in constant length and only a quarter of Nyquist sampled vectors ($14$\farcs$8$ spacing) are plotted for visualisation clarity. Contours are at $N({\rm H}) = 0.2, 0.4, 0.8, 1.2, 2, 2.6$, and $3.2 \times 10^{23}~{\rm cm}^2$ for OMC-3 and $N({\rm H}) = 0.28 \times 10^{23}~{\rm cm}^2$ to $1.48 \times 10^{23}~{\rm cm}^2$, with a step of $0.2 \times 10^{23}~{\rm cm}^2$ for OMC-4. YSOs \citep{fur16} in the observed regions are plotted as blue circles. Prestellar cores \citep{sal15} are plotted as blue triangles. Spatial scale is shown on the bottom right of the figures.
\label{fig:polarization}}
\end{figure*}

In the classical DCF method, the mean field is assumed to be uniform. \citet{hil09} introduced the structure function variant of the DCF method, which allows for a smooth variation in the orientation of the mean field 
(see also \citealp{li22}).
The structure function is defined  as
\beq
\langle\Delta\Phi(\ell)^2\rangle \equiv \frac{1}{N(\ell)} \sum\limits_{i=1}^{N(\ell)} [\Phi(\vecx) - \Phi(\vecx+\vecl)]^2,
\label{eq:sf0}
\eeq
where $\Phi(\vecx)$ is the PA at position $\vecx$,
$\vecl$ is the displacement, and $N(\ell)$ is the number of polarization angle pairs with separation $\ell$. 
When applying the DCF/SF method, it is common to restrict the angle difference, $\Delta \Phi = |\Phi(\vecx) - \Phi(\vecx+\vecl)|$, between any pair of vectors to be in the range $[0,90^\circ]$, although this can lead to errors \citep{li22}. The dispersion in the PAs in the observed OMC-3 and OMC-4 regions under consideration here is small enough that the effect of restricting the angles is negligible. The differences of the fitted $\Delta\Phi_0$ using restriction or not are within the fitting uncertainties. The DCF/SF value of the mean POS field is 
\beq
\bosf = 0.5\sigma_v\sqrt{4\pi\rho(2-\Delta \Phi_0^2)}/\Delta \Phi_0,
\label{eq:sfbo}
\eeq
where $\Delta\Phi_0=\langle\Delta\Phi(\ell\rightarrow 0)^2\rangle^{1/2}$ is in radians. The rms value of the turbulent component of the POS field is
\beq
B_{t\perp} = \bosf \Delta\Phi_0 / \sqrt{2-\Delta \Phi_0^2}.
\label{eq:sfbt}
\eeq
\citet{li22} noted that the approximations made in the derivation of the DCF/SF method are equivalent to assuming that the turbulent field is perpendicular to the mean field.

\subsubsection{Volume density and velocity dispersion}

To apply the DCF method, we need the values of both the volume density and the velocity dispersion. To compute the mean volume density of a filamentary cloud, we assume that the mean depth of the cloud is the same as the mean projected width (or diameter), $D = A / L$, where $A$ is the pixel area occupied by the HAWC+ vectors and $L$ is the length of the cloud. The mean value of volume density is approximated by $n({\rm H}) = N({\rm H})/D$. For dense clumps, we compute the mean radius on the POS by $r = \sqrt{A/\pi}$. We use a diameter of $D = 0.14$ pc for the main cloud to be the depth of the dense clump to compute $n({\rm H})$ in OMC-3. We use $D = 0.30$ pc for the main cloud in OMC-4. The physical sizes of clouds and dense clumps in the two regions are listed in Table \ref{tab:properties}. We do not know exactly the 3D structures of the filamentary clouds and dense clumps. Therefore, the mean volume density estimation could have a large uncertainty. This may actually be the case for the observed OMC-4 region discussed in Section \ref{sec:projection}.

Figure \ref{fig:correlation} shows the 0-moments of the three CO lines with the contours of the Stokes $I$ from dust emission obtained using HAWC+ for the OMC-3 and OMC-4 regions. See Section \ref{subsec:ArchivalData} for detailed information about the archival CO moment maps. The correlation coefficients of the 0-moment and Stokes $I$ maps are 0.28, 0.55, and 0.55, for $^{12}$CO(1-0), $^{13}$CO(1-0), and C$^{18}$O(1-0) respectively in the OMC-3 region. The correlation is poor for $^{12}$CO(1-0) but also not particularly good for $^{13}$CO(1-0) and C$^{18}$O(1-0). In the OMC-4 region, the correlation coefficients are 0.44, 0.74, and 0.84, respectively. The C$^{18}$O(1-0) has the best correlation with HAWC+ Stokes $I$ data. \citet{shi14} used the Nobeyama 45-m telescope, with a resolution of 25\farcs8 ($\sim0.05$ pc), to obtain $^{13}$CO(1-0) and C$^{18}$O(1-0) maps of Orion A. They estimated that the optical depths are $0.05 < \tau_{\rm^{13}CO} < 1.54$ and $0.01 < \tau_{\rm C^{18}O} < 0.18$. Therefore, the optically thin molecular line C$^{18}$O(1-0) is suitable for determining the column density and the velocity dispersion. The velocity dispersion in the observed OMC-3 and OMC-4 regions are in the range of 0.3-0.8 km s$^{-1}$, the corresponding line width, $\Delta v = 2.355\sigma_v$, is about 0.7-1.9 km s$^{-1}$. The channel width of 0.22 km s$^{-1}$ is sufficient to resolve the lines. We discuss the results for OMC-3 and OMC-4 separately below.

\subsubsection{Self-gravity and turbulence}

The DCF method is based on the premise that the turbulent motions are in equipartition with the fluctuations in the magnetic field and that self-gravity does not dominate the dispersion in the orientation of the magnetic field. The importance of self-gravity in a cloud is measured by two dimensionless parameters, the virial parameter, $\avir$, and the mass-to-flux ratio relative to the critical value, $\mu_\Phi$. The virial parameter measures the strength of thermal and turbulent motions relative to that of self-gravity. For ellipsoidal clouds, the virial parameter is $\avir=5\svt^2 r/GM$, where $\svt^2=\sigma_V^2+\cs^2$ is the 1D turbulent plus thermal velocity dispersion and $\cs$ is the isothermal sound speed \citep{ber92}. The conditions for gravitational collapse depend on the shape of the cloud and the internal density distribution, but clouds with $\avir<1$ are generally subject to collapse in the absence of magnetic support. For filamentary clouds, the virial parameter is $\avirf=2\svt^2/G M_\ell$, where $M_\ell$ is the mass per unit length \citep{fie00}.

\begin{figure*}
\includegraphics[angle=0,scale=0.81]{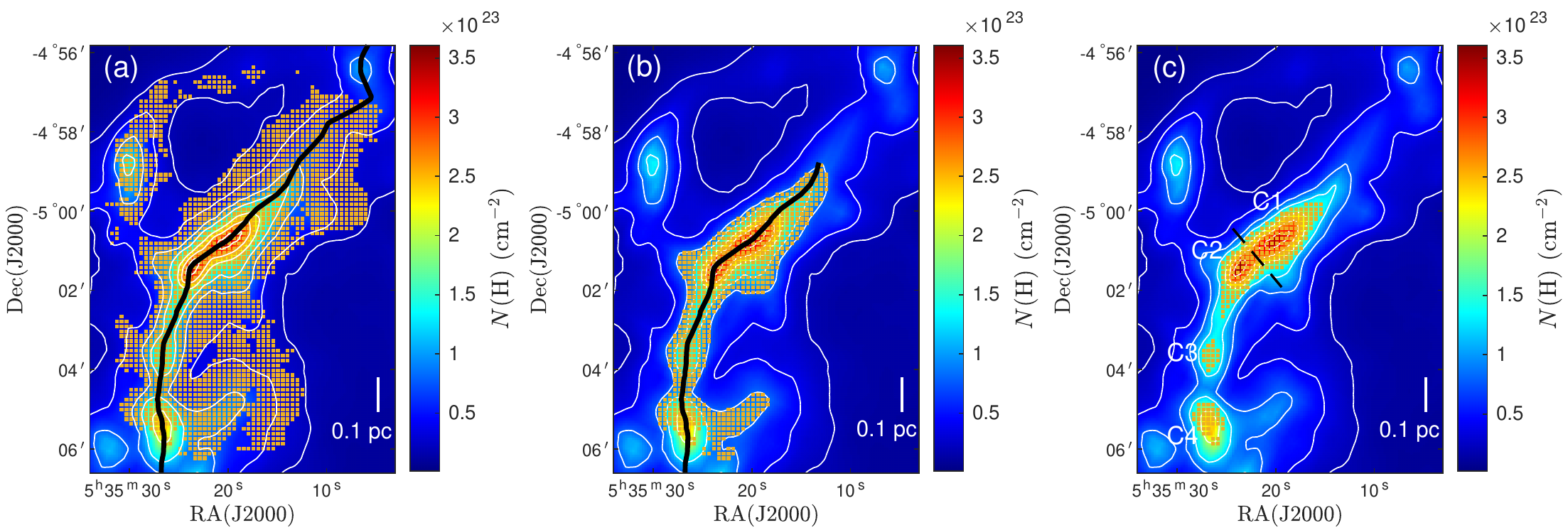}
\caption{(a) Entire map of OMC-3: All pixels (orange squares) that have HAWC+ signal detection as shown in Figure \ref{fig:OMC3_fields} are included in magnetic field estimation. Column density contours are at $N({\rm H}) = 0.2, 0.4, 0.8, 1.2, 2, 2.6$, and $3.2 \times 10^{23} {\rm cm}^2$. (b) Main cloud of OMC-3: Only pixels (orange squares) on the main filamentary cloud with HAWC+ signal detection and column density $N(H) \ge 0.8\times10^{23}$ cm$^{-2}$ are included in the magnetic field strength estimation. (c) Dense clumps of OMC-3: Four dense clumps, labeled from top to bottom as C1 to C4, along the main filamentary cloud with $N(H) \ge 1.4\times10^{23}$ cm$^{-2}$ (orange squares) are identified for magnetic field strength estimation.
\label{fig:OMC-3_dcf}}
\end{figure*}

The relative importance of self-gravity and the magnetic field is measured by the ratio of the mass-to-flux ratio relative to the critical value, $\muphi$.
As shown in the Appendix of \citet{li22}, the value of $\muphi$ is the same for a spherical cloud and a filamentary cloud threaded by a perpendicular magnetic field, $\muphi=2\pi\Sigma/B$, where for a filamentary cloud $\Sigma$ is measured normal to the filament. The value of the magnetic field that enters this relation is the total 3D field.
Using the DCF method, we can measure only the POS value of the normalized mass-to-flux ratio, 
\beq
\muphipos=2\pi N({\rm H})\muh/B_\POS.
\label{eq:muphipos}
\eeq
Here $B_\POS=(B_0^2+B_t^2)^{1/2}=(B_0^2+2B_{t\perp}^2)^{1/2}$ is the total POS field, and the second expression applies if the turbulent field is isotropic (see the discussion below eq. \ref{eq:btp}). If a filamentary cloud is inclined to the plane of the sky by an angle $\gamma_f$, then the actual mass-to-flux ratio is related to the POS value that we can measure by
\beq
\muphi=\muphipos\,\cos\gamma_f\left(\frac{B_\POS}{B_{\rm 3D}}\right)
\label{eq:muphi}
\eeq
\citep{li22}. For a spherical cloud, the same relation holds without the factor $\cos\gamma_f$. 
The median value of $B_\POS/B_{\rm 3D}$ is $\surd 3/2$ if the 3D field is dominated by the uniform component, which would give a median value of $\muphi=0.87\muphipos$ for a spherical cloud. For a filamentary cloud, 
the median value of $\cos\gamma_f$ is also $\surd 3/2$; if $B_\POS/B_{\rm 3D}$ also has its median value, then $\muphi=0.75\muphipos$.

\begin{table*}
\renewcommand*{\thefootnote}{\alph{footnote}}
\caption{Summary of HAWC+ measurement, derived physical properties from \citet{lom14} and \citet{kon18} data, and magnetic field estimation using DCF method for the two regions OMC-3 and OMC-4}
\label{tab:properties}
\begin{tabular}{lccccccc}
\hline
\hline
Region & \multicolumn{7}{c}{OMC-3} \\
\hline
&    & Entire map & Main cloud & \multicolumn{4}{c}{Dense clump} \\
\hline
&   & -- & -- & C1 & C2 & C3 & C4 \\
\hline
& $D$ (pc)\footnotemark & 0.37 & 0.14 & -- & -- & -- & -- \\
& $r$ (pc)\footnotemark & -- & -- & 0.09 & 0.07 & 0.03 & 0.05 \\
& $L$ (pc)\footnotemark & 1.27 & 0.98 & -- & -- & -- & -- \\
& $M~(\msun)$\footnotemark & 662.2 & 290.4 & 68.1 & 47.9 & 6.3 & 16.3 \\
& $M_\ell~(\msun~pc^{-1})$\footnotemark & 521.4 & 296.3 & -- & -- & -- & -- \\
& $N({\rm H})~(\times 10^{23}~{\rm cm}^{-2})$ & $1.27\pm0.91$\footnotemark & $1.86\pm0.82$ & $2.38\pm0.60$ & $2.70\pm0.60$ & $1.51\pm0.07$ & $1.85\pm0.28$ \\
& $n({\rm H})~(\times 10^{5}~{\rm cm}^{-3})$\footnotemark & $1.11\pm0.80$ & $4.25\pm1.87$ & $5.44\pm1.38$ & $6.16\pm1.36$ & $3.45\pm0.16$ & $4.23\pm0.64$ \\
& $\sigma_{\theta}~(^\circ)$ & $25.8\pm0.5$ & $13.7\pm0.4$ & $6.3\pm0.6$ & $13.6\pm1.6$ & $2.7\pm0.6$ & $20.5\pm3.3$ \\
& $\sigma_{V}~({\rm km~s}^{-1})$ & $0.36\pm0.09$ & $0.40\pm0.11$ & $0.43\pm0.06$ & $0.35\pm0.03$ & $0.36\pm0.06$ & $0.24\pm0.04$ \\
& $B_0~({\rm mG})$ & $0.067\pm0.030$ & $0.292\pm0.096$ & $0.775\pm0.143$ & $0.309\pm0.047$ & $1.21\pm0.21$ & $0.112\pm0.023$ \\
& $B_{t\perp}~({\rm mG})$\footnotemark & $0.033\pm0.015$ & $0.071\pm0.024$ & $0.086\pm0.019$ & $0.074\pm0.017$ & $0.057\pm0.022$ & $0.042\pm0.014$ \\
& $\ma/{\rm cos} \gamma$\footnotemark & $1.67\pm1.05$ & $0.84\pm0.37$ & $0.38\pm0.10$ & $0.84\pm0.18$ & $0.16\pm0.04$ & $1.30\pm0.35$ \\
& $\avir,\,\avirf$\footnotemark & $0.16\pm0.08$ & $0.33\pm0.19$ & $0.36\pm0.13$ & $0.30\pm0.09$ & $1.13\pm0.39$ & $0.38\pm0.14$ \\
& $\muphipos$\footnotemark & $5.1\pm4.2$ & $2.2\pm1.2$ & $1.1\pm0.4$ & $3.0\pm0.8$ & $0.5\pm0.1$ & $5.0\pm1.2$ \\
& $M_\ell/M_{\ell,{\rm crit}}$\footnotemark & $4.0\pm2.1$ & $1.8\pm0.7$ & -- & -- & -- & -- \\
\hline
\hline
Region & \multicolumn{7}{c}{OMC-4} \\
\hline
&    & Entire map & Main cloud & \multicolumn{4}{c}{Dense clump} \\
\hline
&   & -- & -- & C1 & C2 & C3 & C4 \\
\hline
& $D$ (pc) & 0.31 & 0.30 & -- & -- & -- & -- \\
& $r$ (pc) & -- & -- & 0.06 & 0.03 & 0.09 & 0.11 \\
& $L$ (pc) & 1.21 & 0.75 & -- & -- & -- & -- \\
& $M~(\msun)$ & 272.8 & 197.0 & 9.2 & 2.2 & 23.1 & 43.6 \\
& $M_\ell~(\msun~pc^{-1})$ & 225.5 & 261.4 & -- & -- & -- & -- \\
& $N({\rm H})~(\times 10^{23}~{\rm cm}^{-2})$ & $0.64\pm0.26$ & $0.77\pm0.24$ & $0.76\pm0.04$ & $0.70\pm0.01$ & $0.79\pm0.07$ & $1.01\pm0.23$ \\
& $n({\rm H})~(\times 10^{5}~{\rm cm}^{-3})$ & $0.67\pm0.27$ & $0.82\pm0.25$ & $0.81\pm0.04$ & $0.74\pm0.01$ & $0.84\pm0.07$ & $1.08\pm0.25$ \\
& $\sigma_{\theta}~(^\circ)$ & $14.8\pm0.3$ & $13.6\pm0.4$ & $5.5\pm0.8$ & $4.3\pm1.1$ & $8.8\pm0.8$ & $11.5\pm0.8$ \\
& $\sigma_{V}~({\rm km~s}^{-1})$ & $0.49\pm0.17$ & $0.53\pm0.17$ & $0.80\pm0.04$ & $0.55\pm0.05$ & $0.72\pm0.07$ & $0.45\pm0.11$ \\
& $B_0~({\rm mG})$ & $0.129\pm0.053$ & $0.171\pm0.066$ & $0.644\pm0.036$ & $0.547\pm0.053$ & $0.367\pm0.038$ & $0.199\pm0.052$ \\
& $B_{t\perp}~({\rm mG})$ & $0.034\pm0.014$ & $0.041\pm0.016$ & $0.062\pm0.013$ & $0.041\pm0.017$ & $0.057\pm0.009$ & $0.040\pm0.011$ \\
& $\ma/{\rm cos} \gamma$ & $0.92\pm0.53$ & $0.84\pm0.44$ & $0.33\pm0.03$ & $0.26\pm0.03$ & $0.54\pm0.08$ & $0.71\pm0.26$ \\
& $\avir,\,\avirf$ & $0.59\pm0.48$ & $0.59\pm0.42$ & $5.1\pm0.5$ & $5.53\pm1.01$ & $2.63\pm0.54$ & $0.76\pm0.39$ \\
& $\muphipos$ & $1.7\pm0.9$ & $1.5\pm0.7$ & $0.4\pm0.0$ & $0.5\pm0.1$ & $0.8\pm0.1$ & $1.9\pm0.6$ \\
& $M_\ell/M_{\ell,{\rm crit}}$ & $1.2\pm0.6$ & $1.1\pm0.5$ & -- & -- & -- & -- \\
\hline
\hline
\end{tabular}
\begin{flushleft}
\fnt{1} {$^{\rm a}$} The mean projected width, $D = {\rm Area} / L$, of the gas cloud when it is approximated as a filamentary cloud. The Area is the area covered by the pixels shown in Figures \ref{fig:OMC-3_dcf} and \ref{fig:OMC-4_dcf}.\\
\fnt{2} {$^{\rm b}$} The mean projected radius, $r = \sqrt{({\rm Area}/\pi)}$, of the gas cloud when it is approximated as a spherical cloud. \\
\fnt{3} {$^{\rm c}$} The length of the cloud when it is approximated as a filamentary cloud.\\
\fnt{4} {$^{\rm d}$} The mass of the cloud.\\
\fnt{5} {$^{\rm e}$} The mass per unit length, $M_\ell=M/L$, of a filamentary cloud.\\
\fnt{6} {$^{\rm f}$} The dispersions of $N({\rm H})$ and $\sigma_{V}$ are the weighted standard deviation using polarization angle measurements error for weighting. The uncertainty for $\sigma_{\theta}$ is estimated as $\sigma_{\theta}/\sqrt{N_p}$, where $N_p$ is the number of polarization measurements \citep[cf.][]{pat21}. The uncertainties of other derived results are obtained from the propagation of uncertainties. \\
\fnt{7} {$^{\rm g}$} For filamentary clouds, mean volume density, $n = N({\rm H})/D$, where the LOS depth of the cloud is assumed to be the mean projected width $D$. It is the same for dense clumps but using the LOS depth of the main cloud.\\
\fnt{8} {$^{\rm h}$} $B_{t\perp} = B_0~{\rm tan}(\sigma_\theta)$ (eq. \ref{eq:btp}).\\
\fnt{9} {$^{\rm i}$} 3D \alfven Mach number, $\surd 3\sigma_V/(B_0/\sqrt{4\pi\rho})$, based on the mean POS field, $B_0$, and assuming isotropic turbulence. $\gamma$ is the inclination angle between the 3D mean field direction and the POS.\\
\fnt{10} {$^{\rm j}$} For filamentary clouds (the entire maps of OMC-3 and OMC-4 and the main cloud of region OMC-3), the virial parameter is $\avirf = 2\sigma_{V,\,\rm tot}^2L/(GM)$. For spherical clouds, such as the main cloud of OMC-4 and all the dense clumps, $\avir = 5\sigma_{V,\,\rm tot}^2r/(GM)$. The velocity dispersion, $\sigma_{V,\,\rm tot}$, includes the thermal component of gas with an assumed temperature of 15 K. \\
\fnt{11} {$^{\rm k}$} For filamentary clouds (the entire maps of OMC-3 and OMC-4 and the main cloud of region OMC-3), the POS mass-to-flux ratio in units of the critical value is based on the total magnetic field strength $\sqrt{B_0^2+B_t^2}$ estimated on the POS. For clump-like cloud, such as the main cloud of OMC-4 and all the dense clumps, $\muphi = 7.6\times10^{-21} N({\rm H}_2)/\sqrt{B_0^2+B_t^2}$, where $B_0$ and $B_t$ are in $\mmug$. \\
\fnt{12} {$^{\rm l}$} The ratio of length mass to the critical value for filamentary clouds, $M_\ell/M_{\crit,\ell} = \left(\muphipos^{-2}+\avirf^2\right)^{-1/2}$, using the $\muphipos$ on POS.
\end{flushleft}
\end{table*}

\begin{table}
\caption{Comparison of magnetic field strength estimations using DCF and DCF/SF methods for OMC-3 and OMC-4 regions}
\label{tab:properties_sf}
\renewcommand*{\thefootnote}{\alph{footnote}}
\begin{tabular}{lcc}
\hline
\hline
Region & \multicolumn{2}{c}{OMC-3} \\
\hline
DCF/SF & entire map & main cloud \\
\hline
$\Delta \Phi_0~(^\circ)$ & $33.3\pm1.6$ & $18.2\pm0.6$ \\
$\bosf~({\rm mG})$\footnotemark & $0.072\pm0.019$ & $0.312\pm0.056$ \\
$B_{t\perp}~({\rm mG})$\footnotemark & $0.033\pm0.009$ & $0.071\pm0.013$ \\
\hline
DCF & & \\
\hline
$B_0~({\rm mG})$ & $0.067\pm0.030$ & $0.292\pm0.096$ \\
$B_{t\perp}~({\rm mG})$ & $0.033\pm0.015$ & $0.071\pm0.024$ \\
\hline
\hline
Region & \multicolumn{2}{c}{OMC-4} \\
\hline
DCF/SF & entire map & main cloud \\
\hline
$\Delta \Phi_0~(^\circ)$ & $15.5\pm0.5$ & $18.4\pm0.5$ \\
$\bosf~({\rm mG})$ & $0.175\pm0.062$ & $0.178\pm0.058$ \\
$B_{t\perp}~({\rm mG})$ & $0.034\pm0.012$ & $0.042\pm0.013$ \\
\hline
DCF & & \\
\hline
$B_0~({\rm mG})$ & $0.129\pm0.053$ & $0.171\pm0.066$ \\
$B_{t\perp}~({\rm mG})$ & $0.034\pm0.014$ & $0.041\pm0.016$ \\
\hline
\hline
\end{tabular}
\begin{flushleft}
\fnt{1} {$^{\rm a}$}
Magnitude of the spatially varying mean field (eq. \ref{eq:sfbo}).\\
\fnt{2} {$^{\rm b}$} Turbulent component of the POS field strength normal to the mean field (eq. \ref{eq:sfbt}).\\
\end{flushleft}
\end{table}

\begin{figure*}
\includegraphics[angle=0,scale=0.8]{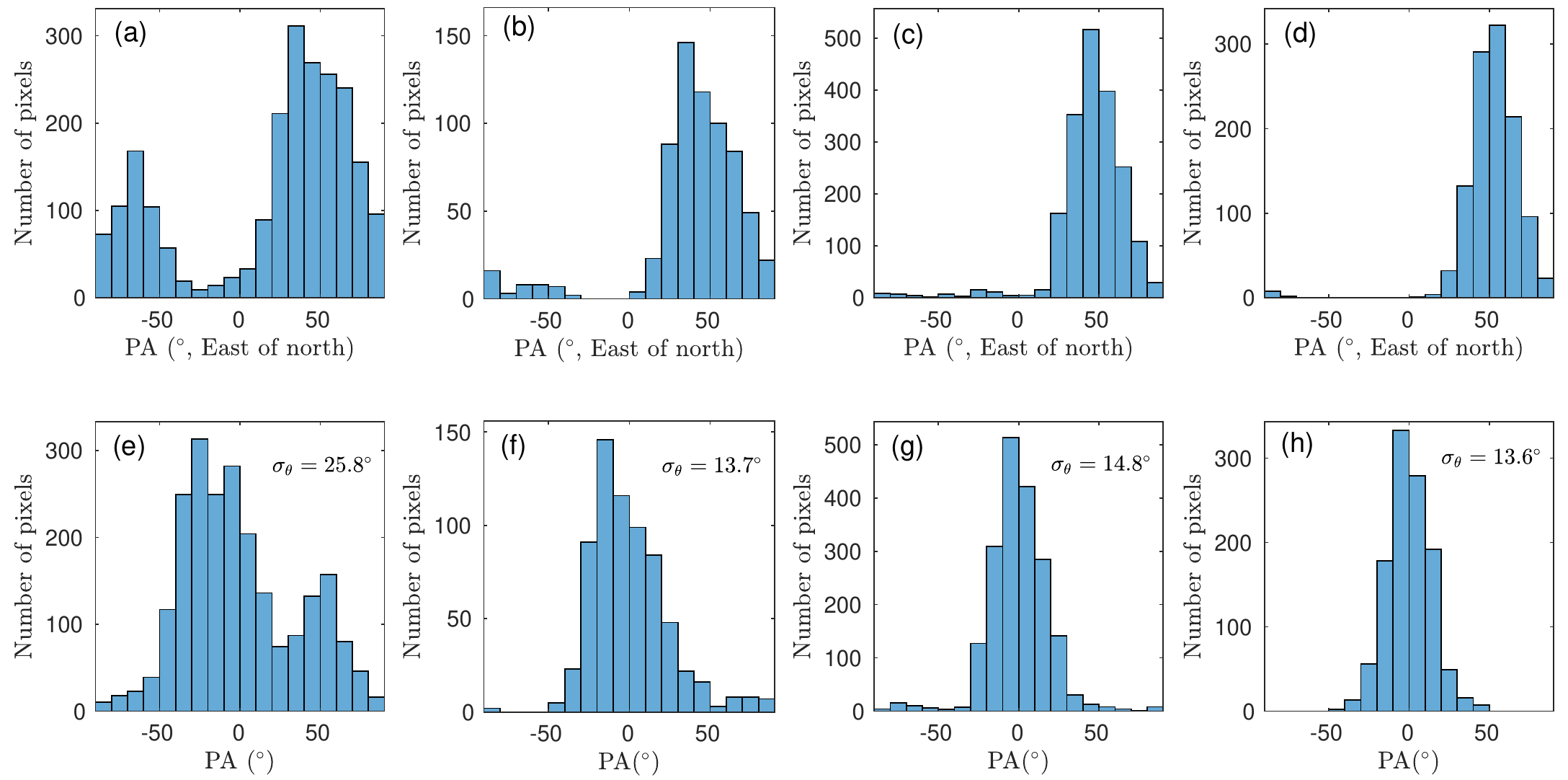}
\caption{Top four panels are the PA distributions of (a) Entire map of OMC-3 region, (b) main cloud of OMC-3 region, (c) entire map of OMC-4 region, and (d) main cloud of OMC-4. The bottom four panels are the corresponding PA distributions after shifting the PAs so that the dispersions of angles are minimum. The error-weighted angle dispersions are marked in the bottom panels.
\label{fig:PA}}
\end{figure*}

\subsection{Results for OMC-3}

\label{sec:OMC-3_dcf}

In order to better visualise the inferred magnetic field orientation, we replot the polarization measurements in Figures \ref{fig:OMC3_fields} and \ref{fig:OMC4_fields} but with constant length on top of the column density map from \citet{lom14} in Figure \ref{fig:polarization}. We can see a well-defined main filamentary cloud outlined with some other filamentary-like gas structures at the upper-left and lower-right sides of the cloud. We want to estimate the magnetic field strengths on different size scales in this region, (1) the entire map, (2) the main cloud, and (3) the dense clumps in the main cloud. 

The mean values and derived physical properties of these three regions are given in Table \ref{tab:properties}. The entire map includes all the HAWC+ detections, shown in Figure \ref{fig:polarization} as marked by the orange squares in Figure \ref{fig:OMC-3_dcf}a. The threshold column density, $0.2\times 10^{23}$~cm\ee, is much less than the mean column density, $1.27\times 10^{23}$~cm\ee, because the column density in each pixel is weighted by the error in the polarization measurement and the error is smaller at higher column densities. The black curve shows the location of the peak column density at each declination; it has a length of 1.27 pc. The column density thresholds for the main cloud ($N({\rm H}) = 0.8\times10^{23}$~cm\ee) and the dense clumps ($N({\rm H}) = 1.4\times10^{23}$~cm\ee) were chosen to study the physical conditions of the central main filamentary cloud and the dense clumps at different density and size scales. The main filamentary cloud is shown in Figure \ref{fig:OMC-3_dcf}b; the black curve showing the location of the peak column density has a length of 0.98 pc. Four dense clumps along the main filamentary clouds are shown in Figure \ref{fig:OMC-3_dcf}c. In both cases, only the polarization measurements at pixels above the density threshold are included in computing the dispersion in polarization angles.

The PA distributions for the entire map and the main cloud in the OMC-3 region are shown in Figure \ref{fig:PA}a and b. The two groups peak at about $-65^\circ$ and $35^\circ$ East of North, about $100^\circ$ apart from each other. When using the DCF method, we shift the PAs so that the angular dispersion is a minimum \citep{pad01}. The corresponding shifted PA distributions are shown in Figure \ref{fig:PA}e and \ref{fig:PA}f, respectively. In Figure \ref{fig:PA}e, we can see a two distinct groups of PAs but not in Figure \ref{fig:PA}f, in which only pixels with $N({\rm H}) \ge 0.8\times10^{23} {\rm cm}^{-2}$ are plotted. This means that the smaller group of PAs (centered at $\theta=55^\circ$ in Figure \ref{fig:PA}e) comes mostly from pixels with lower column densities. Without the PAs from low column-density regions, the dispersion of the PAs, $\sigma_\theta = 13.7^\circ$, in the main cloud is smaller than that in the whole region of $25.8^\circ$, as indicated in the figure.

The estimated field strengths in each region using the standard DCF method are presented in Table \ref{tab:properties} and are compared with those from DCF/SF method in Table \ref{tab:properties_sf}. We do not apply the DCF/SF method to individual dense clumps due to the small numbers of polarization vectors in the dense clumps precludes statistically reliable fitting. From Table \ref{tab:properties_sf}, we can see the field strength estimated using the DCF/SF method is slightly larger than that using the DCF method. The reason is straightforward: In the classical DCF method, the total dispersion of the polarization angles $\sigma_{\theta}$ is used to determine the field strength of the mean field, $B_0$, whereas in the DCF/SF method, the larger-scale variation in the direction of the mean field is excluded from the dispersion. Because the two estimates of the field strength are close, we shall use only one, the DCF estimate, in the following discussion of the physical properties of the cloud structures.

The measured and derived physical properties of the structures in OMC-3 are listed in Table \ref{tab:properties}. In the case of entire map of OMC-3, there are more structures than a simple single filament cloud assumed in the field strength estimation. HAWC+ is basically looking at a region with at least two different field structures associated with several filamentary structures. The field strength estimated for the entire region, $\sim 67$ \mug, is much lower than the main cloud and the dense clumps. Also, because the measurement error of PA is slightly smaller for higher column density pixels, the mean column density $N({\rm H})$ in Table \ref{tab:properties} is generally smaller than the mean column density without weighting. The differences are a few percents for the dense clumps to about 20 per cent for the main cloud. The largest difference is in the case of entire region of OMC-3, about 40 per cent. Without weighting, the total mass in the case of the entire map of OMC-3 is about 403 $\msun$. With a smaller mass, $\avirf$ will be larger and $\muphipos$ is smaller correspondingly. However, this does not change the conclusion that the entire region is subvirial and magnetically supercritical.

The mass per unit length, $M_\ell$, of the main filamentary cloud is estimated to be $296~\msun$ pc$^{-1}$. \citet{sch21} found the average $M_\ell \sim 200~\msun {\rm pc}^{-1}$. The main reason of the difference between our value and theirs is due to the error measurement weighting that we adopted in computing the mean column density. Using $\avirf$ and $\muphipos$, $M_\ell/M_{\crit,\ell} = \left(\muphi^{-2}+\avirf^2\right)^{-1/2} \sim 1.3$, if $\cos\gamma_f = 1$. The main cloud is expected to be unstable with fragmentation. In fact, there are prestellar cores \citep{sal15} and many young stellar objects (YSOs) \citep{fur16} found in the main cloud (see Figure \ref{fig:polarization}). Dense clump C1 is about magnetically critical and only the small dense clump C3 is clearly magnetically subcritical. All the other gas structures are magnetically supercritical. Provided that the field is not too far from the plane of the sky ($\gamma\la 45^\circ$), the majority of the gas structures in OMC-3 are trans-\alfvenic and sub-\alfvenicstop. The average field strength of the dense clumps is about 0.6~mG, albeit with a large dispersion of 0.49 mG that larger than the uncertainties of field strengths of dense clumps. We conclude that all the dense clumps are subvirial, which indicates that gravity is dominant in the OMC-3 region. Turbulence and magnetic forces will not be able to prevent the gas structures from undergoing gravitational collapse to form dense cores and stars. 

Morphologically, the inferred magnetic field orientation is mostly perpendicular to the main filamentary cloud, except when the cloud turns almost south near the dense clump at the bottom of the map. For the main filamentary cloud from near the top-right corner of the map down to the dense clump C3, the mean of the angle differences between the cloud axis and the polarization PAs within 0.075 pc from the axis is $86.8^\circ \pm 14.7^\circ$. However, when we look at the low column density gas around the main cloud, the fields appear to have a large change in orientation. The angle difference is about $100^\circ$ as shown in Figure \ref{fig:PA}e. It appears that there is a magnetic field structure associated with the dense filamentary cloud and another magnetic field structure associated with lower density gas as a background. The field has small distortions at dense clumps C2 and C4. Unfortunately, the dense clump C4 is close to the edge of the FOV of HAWC+ and we do not have complete coverage of polarization of the entire dense clump. Therefore, the estimation for dense clump C4 will have a larger uncertainty. Since these two dense clumps are trans-\alfvenic based on the LOS velocity dispersion and the POS field strength estimate, the larger distortion in the magnetic fields (i.e. larger $\sigma_\theta$) of these two dense clumps is probably due to self-gravity. Recent analysis by \citet{liub21} using ALMA and JVLA data suggest that the magnetic field in the inner 100 au of OMC-3/MMS6, which corresponds to the dense clump C2, has a magnetic field pointing along the main filament cloud. If so, that corresponds to about $90^\circ$ change in the magnetic field orientation from a scale $\sim 0.1$ pc to a scale $\sim 10^{-4}$ pc. High resolution polarization mapping of these two dense clumps is needed to determine how the magnetic field changes below 0.1 pc size scale.

\begin{figure}
\includegraphics[angle=0,scale=0.8]{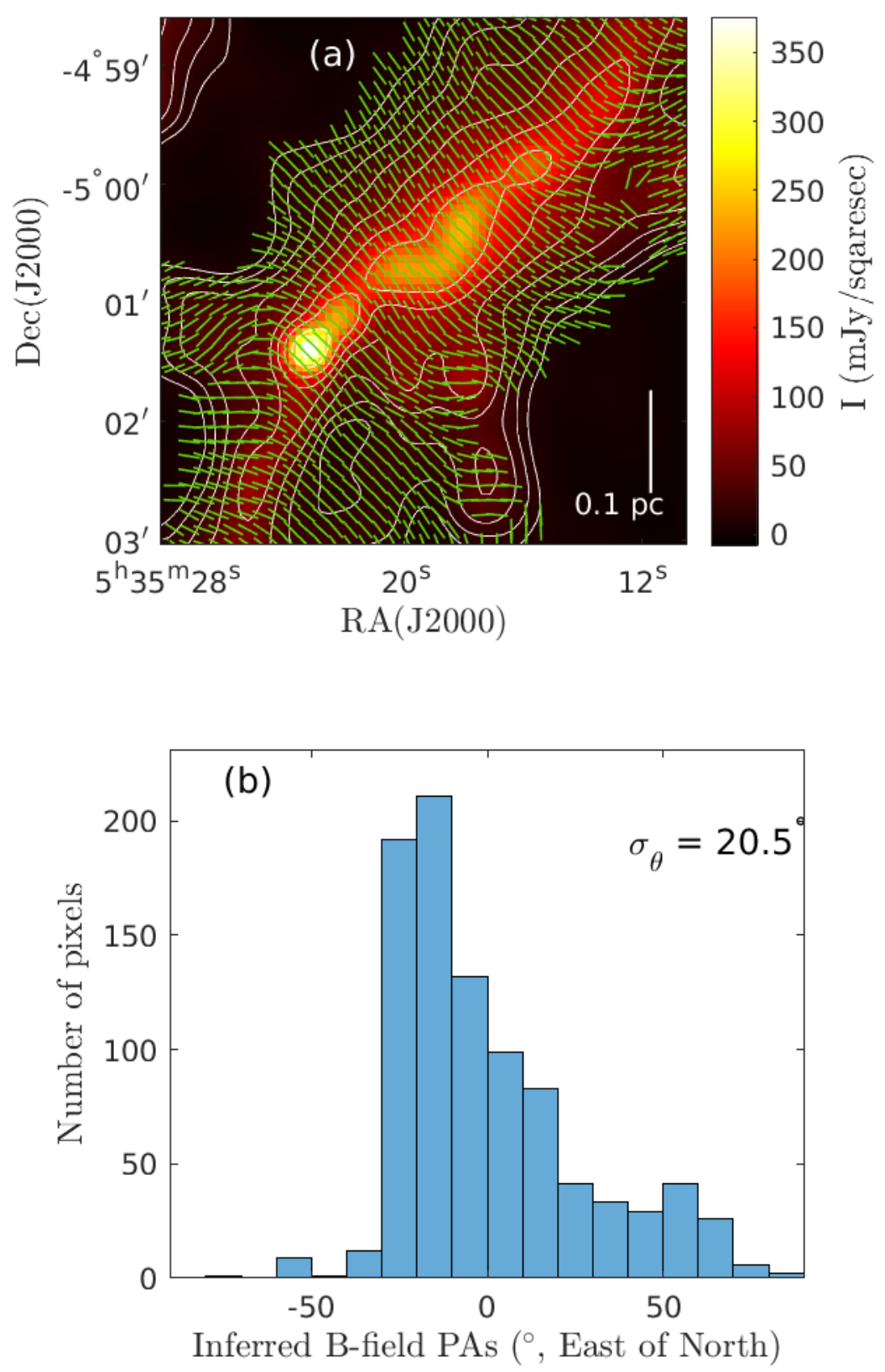}
\caption{(a) Same as the Figure \ref{fig:OMC3_fields} but shows only the region as the right panel of the figure 1 in \citet{zie22} for comparison. The inferred magnetic field orientations are shown in constant length for visualisation purposes. (b) The PA distribution of inferred magnetic field vectors in the region shown in (a).  The error-weighted dispersion of PAs is marked in the figure.
\label{fig:ziecomp}}
\end{figure}

\subsubsection{Comparison with previous results}

Our estimate of the field strength in dense clump C2, $B_0=0.31$~mG, is somewhat less than the 0.64~mG obtained by \citet{mat05} for the same dense clump (labeled MMS6) using much less polarization measurements. \citet{poi10} found a mean field strength of MMS1 to MMS7, corresponding to the main cloud in our study, to be 0.19 mG using the DCF/SF method. Their estimate of the field strength is lower than our estimations using either the standard DCF method or the DCF/SF method (see table \ref{tab:properties_sf}).

\begin{figure*}
\includegraphics[angle=0,scale=0.81]{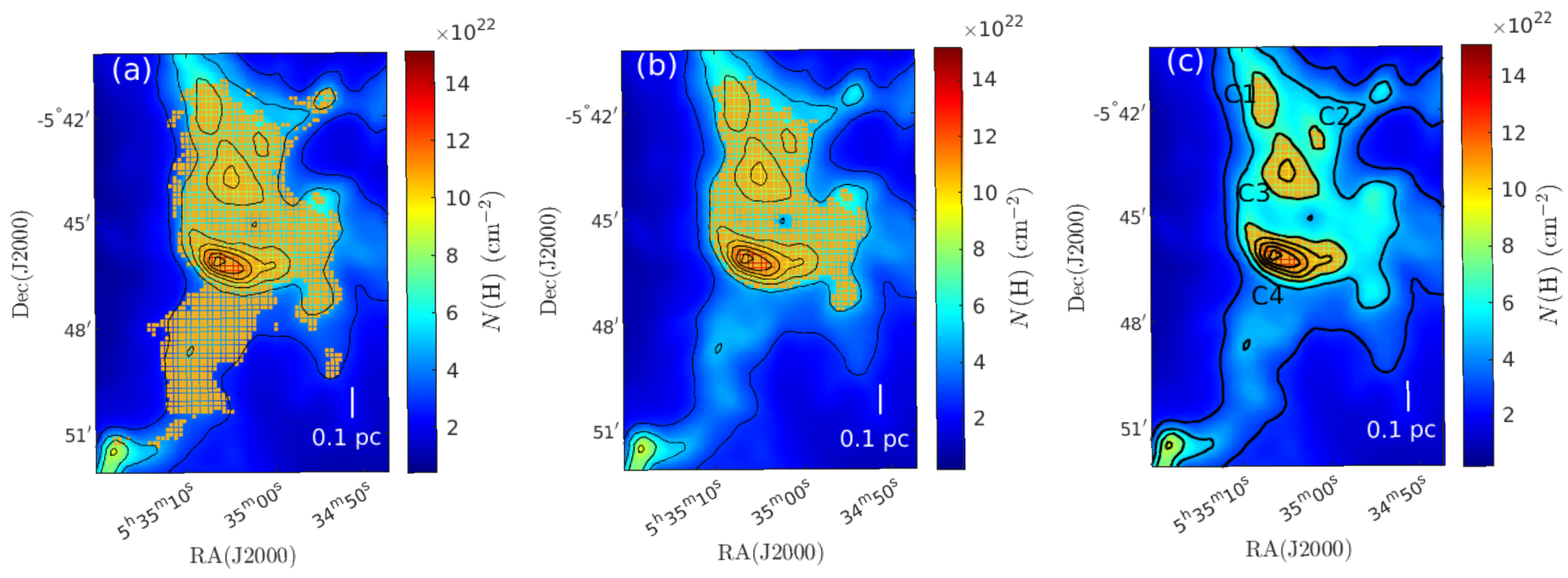}
\caption{(a) Entire map of OMC-4: All pixels (orange squares) that have HAWC+ signal detection are included in magnetic field estimation. Contours are at column density $N({\rm H})$ from 0.28 to $1.48\times10^{23}$ cm$^{-2}$ with a step of $0.2\times10^{23}$ cm$^{-2}$. (b) Main cloud of OMC-4: Only pixels (orange squares) in the region with HAWC+ signal detection and column density $N(H) \ge 0.5\times10^{23}$ cm$^{-2}$ are included in the magnetic field strength estimation in this case. (c) Dense clumps of OMC-4: Four dense clumps, labeled from top to bottom as C1 to C4 with $N(H) \ge 0.68\times10^{23}$ cm$^{-2}$ (orange squares) are identified for magnetic field strength estimation.
\label{fig:OMC-4_dcf}}
\end{figure*}

\citet{zie22} recently reported the polarization results of OMC-3 using HAWC+ in Band D and Band E during cycle 7. We only have Band E data and therefore our discussion is based only on their Band E results. They used the chop-nod observing mode and their field of view is smaller than ours in Band E. To compare with their results, we extract the same FOV region as shown in the right panel of their figure 1 and plot our map in Figure \ref{fig:ziecomp}. Their plots, using E-vectors, indicates that the inferred magnetic field orientation is almost perpendicular to the filamentary cloud. It is consistent with our polarization map in Figure \ref{fig:ziecomp}a. The histogram of PAs in this region is shown in Figure \ref{fig:ziecomp}b. Our PA distribution is slightly different from theirs, which is shown in the bottom panel of their figure 2. Visually, we can see that our map (Figure \ref{fig:ziecomp}a) has more polarization measurements that have larger angular dispersion near the left and bottom edges of the map. This is reflected in the small difference in the shape of our PA histogram and theirs. However, in their magnetic field estimation, they only use PAs inside the 1$\sigma$ region around the mean PA to compute the angle dispersion. This difference produces that their angle dispersion is only $8.68^\circ$, less than half of the dispersion, $20.5^\circ$, shown in our map (Figure \ref{fig:ziecomp}b). Thus, our estimated magnetic field strength for this smaller region is 128.7 $\mmug$, about 63 per cent of the estimated value of 205.4 $\mmug$ by \citet{zie22}.

\subsection{Results for OMC-4}
\label{sec:OMC-4_dcf}

The gas distribution of the observed region by HAWC+ in OMC-4 is very different from that in OMC-3. Instead of any well-defined filamentary clouds as in OMC-3, we see the observed OMC-4 region is clumpy with some possible filamentary-like connections among the dense clumps. One elongated gas structure appears like a hook. In the right panel of Figure \ref{fig:polarization}, we plot the inferred magnetic field orientation with constant length from HAWC+ over the column density map for the observed OMC-4 region. Similar to OMC-3, we want to find out the magnetic field strengths at different size scales in OMC-4.

We again use three different column densities to restrict the subregions to be examined, as shown in Figure \ref{fig:OMC-4_dcf}. For the entire map, we approximate the observed OMC-4 region as a thick filamentary cloud running from top to bottom. For the dense main cloud, by considering pixels only have column density $N({\rm H}) \ge 0.5 \times 10^{23} {\rm cm}^{-2}$. We still approximate the gas cloud as a filamentary cloud similar to the entire map. For dense clumps, we consider the pixels with $N({\rm H}) \ge 0.68 \times 10^{23} {\rm cm}^{-2}$. There are four dense clumps identified. Three out of four dense clumps have lower column density than those in OMC-3 region. If we use the same column density threshold for dense clumps in OMC-3, we shall have only one dense clump in the observed OMC-4 region. 

When looking at a larger scale, the gas structures of Orion A basically runs from North to South roughly vertically \citep[see figures in e.g.][]{kon18}. In OMC-4, the gas distribution is more complex and it is possible that several low density filamentary and clumpy structures overlap or entangle in this region. The observed OMC-4 region by HAWC+ is only a portion of OMC-4. From \citet{kon18} C$^{18}$O(1-0) data, the velocity dispersion of the dense clumps in OMC-4 is about double of those in OMC-3 as shown in Table \ref{tab:properties}. However, HAWC+ shows that the inferred magnetic field in Figure \ref{fig:polarization} is even more uniform than that in the OMC-3 region. Figure \ref{fig:PA}c shows the PA distribution of the observed region OMC-4 and the distribution of the shifted PAs is shown in Figure \ref{fig:PA}g. The error-weighted dispersion of polarization angles in OMC-4 region is only $14.8^\circ$.

\begin{figure*}
\includegraphics[angle=0,scale=0.62]{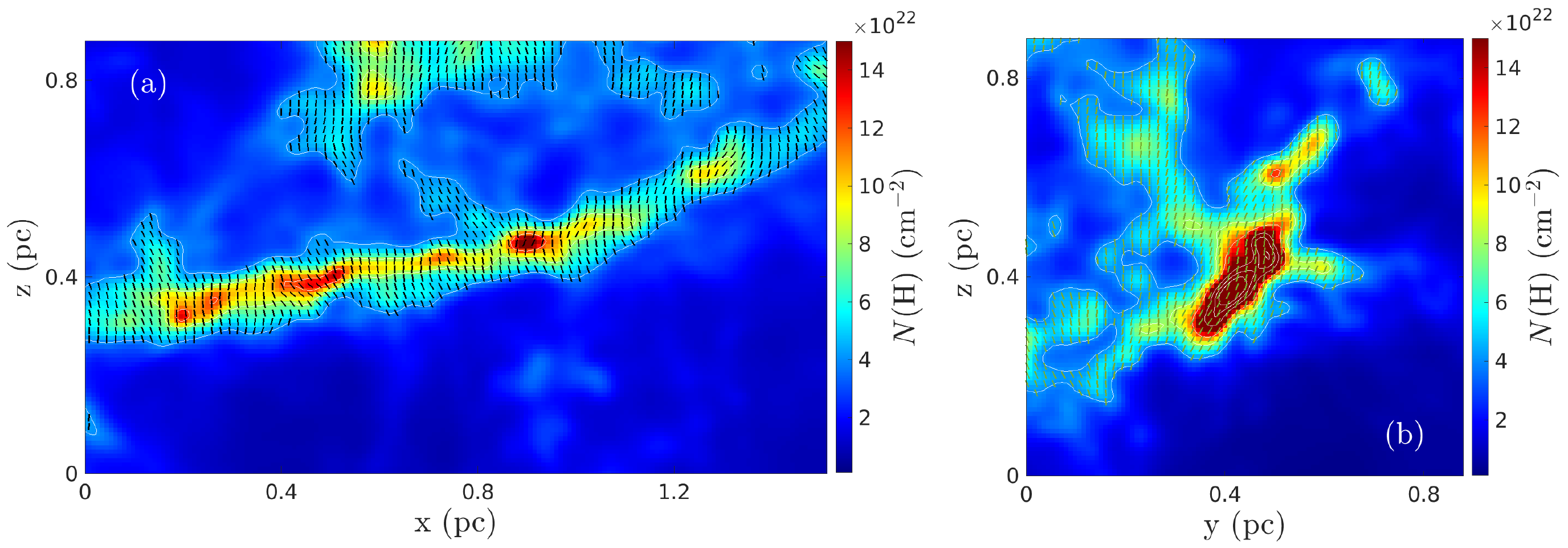}
\caption{(a) The column density map of a simulated filamentary cloud. The viewing direction is along the y-axis. The black lines show the magnetic field mostly perpendicular to the filamentary cloud long axis. The first contour is at $N({\rm H}) = 0.4\times10^{23} {\rm cm}^{-2}$. Only the magnetic field lines at those pixels within the first contour are plotted. (b) Same as panel (a) but the viewing direction is along x-axis, a $90^{\circ}$ counter clockwise rotation about the z-axis from the panel (a). The elongated dense structure near the middle is the projected image of the filamentary cloud in panel (a). Magnetic field lines are still quit uniform but pointing close to the long axis of the elongated gas structure, resembling similar field orientation appearance of the dense clump C4 in Figure \ref{fig:polarization}.
\label{fig:simcloud}}
\end{figure*}

The measured and derived physical properties of the gas structures in OMC-4 region are listed in Table \ref{tab:properties}. We estimate the physical properties of the gas structures in three different size scales, as shown in Figure \ref{fig:OMC-4_dcf}. In the case of the entire map, the mean projected width of the cloud is 0.31 pc using all the pixels marked in Figure \ref{fig:OMC-4_dcf}a. By treating the entire map as a filamentary structure, the $M_\ell/M_{\crit,\ell} \sim 1.2$, if $\cos\gamma_f = 1$. The region is slightly magnetically supercritical and gravitationally stable. We note that in the OMC-4 region, the difference between the mean column densities computed with and without the measurement error weighting is less than 12 per cent, smaller than that in OMC-3 region. The physical properties of the denser main cloud are similar to the entire map. For the four dense clumps, only C4 is magnetically supercritical. The others are highly magnetically subcritical. The reason that the main gas cloud is supercritical is mainly from the contribution of dense clump C4 in computing $\muphipos$. Since the $\avir$ of the three clumps C1 to C3 are much higher than unity, the influence from gravity is small in this region. Magnetic field is dynamically dominating the turbulence and gravity except in the elongated dense clump C4. The average field strength of the dense clumps in OMC-4 is about 0.44 mG, a little smaller than the average field strength of the dense clumps in OMC-3 region. We cannot find magnetic field strength estimation on OMC-4 by others to compare with. The estimated field strengths using the DCF/SF method are close to the values using DCF method and are listed in Table \ref{tab:properties_sf}. We also do not apply DCF/SF method for dense clumps in this region.

The average $\sigma_V$ of dense clumps in OMC-3 and OMC-4 are 0.35 km s$^{-1}$ and 0.63 km s$^{-1}$, respectively. They are different by about a factor of two. If the linewidth-size relation $\sigma_{\rm NT} \propto \ell^{1/2}$ holds at this size scale, the size of OMC-4 dense clumps are expected to be about 3.2 times that of the dense clumps in OMC-3. However, the average projected radius of the dense clumps in OMC-3 region is about the same as those in OMC-4 region. This implies that the LOS depths of dense clumps in OMC-4 regions could be significantly larger than the projected diameters in our assumption of spherical dense clumps in OMC-4. Is it possible that the linewidth-size relation is significantly different for the dense clumps in Orion A? \citet{cas95} point out that massive cloud cores in Orion A and B have a linewidth-size relation $\sigma_{\rm NT} \propto \ell^q$ with $q = 0.21\pm0.03$, significantly smaller than $0.53\pm0.07$ for low mass cloud cores. This does not solve the large discrepancy seen among the dense clumps in OMC-3 and OMC-4 regions because a smaller $q$ means an even larger LOS depth for dense clumps in OMC-4. A possible explanation is that in the observed OMC-4 region, the dense clumps are actually a number of filamentary structures lying close to the LOS. When viewing close to the long axis, a not very dense filament will appear like a elongated dense clump similar to the dense clump C4 in Figure \ref{fig:OMC-4_dcf}c. The projected column density will be higher than if it is a sphere at the same volume density. The other three lower density clumps could also be compose of several low density filamentary substructures. The larger-than-expected velocity dispersion in the observed OMC-4 region may be explained if our view is close to the long axes of several filamentary substructures. This interpretation will be discussed in the following section.

\subsubsection{Projection effect on interpreting the physical states of OMC-4}
\label{sec:projection}

To demonstrate that projection effect can substantially change the angle between the orientation of magnetic field with respect to a filamentary cloud long axis on the POS, we use the simulation data from the infrared filamentary dark cloud simulation by \citet{li19}. The isothermal ideal MHD simulation is a driven turbulence simulation with gravity. The system is maintained at thermal Mach number of 10 all the time after gravity is turned on, with an initial Alfven Mach number of 1 and virial parameter of 1. A long filamentary cloud forms and remains intact to the end of the simulation about 900,000 years after gravity is turned on. The large-scale magnetic field near the filamentary cloud is about perpendicular to the cloud long axis. The details on the simulation methods and results can be found in \citet{li19}. Here we use a snapshot data from the simulation at about $4.2\times10^5$ years after gravity is turned on for comparison. Figure \ref{fig:simcloud}a shows the column density map of the filamentary cloud with the magnetic field orientations. We use this data snapshot because the column density of the cloud is similar to the observed region of OMC-4. The filamentary cloud becomes denser in time from gravitational collapse. The viewing direction is along the y-axis. The inferred magnetic field vectors on the projection plane are created using the definition of Stokes parameters in \citet{zwe96} based on the 3D components of magnetic field and density in the simulation
\begin{align}
Q+iU &=1/N \int f(y)\frac{(B_x+iB_z)^2}{B_x^2+B_z^2}{\rm cos}^2\gamma dy, \\
\phi &=\frac{1}{2}{\rm arctan}\frac{U}{Q}.
\end{align}
Here $N=\int f(y)dy$ is the column density, \citet{zwe96} assumed cos$^2 \gamma$ to be 1, and $\phi$ is the polarization angle. Grains are assumed to be perfectly aligned with magnetic field.

Only vectors in the pixels that have column density $N({\rm H}) \ge 0.4\times 10^{23} {\rm cm}^{-2}$ are shown in the figure for visual clarity. The magnetic field on the POS is closely perpendicular to the filamentary cloud long axis, with an angle of $83.9^\circ \pm 15.4^\circ$ between the axis of filamentary cloud and the mean field vectors shown in Figure \ref{fig:simcloud}a. In Figure \ref{fig:simcloud}b, the viewing angle is rotated by $90^\circ$ about the vertical axis that we are now looking along the x-axis. In this orientation, the long filamentary cloud axis is making an inclination angle of about $23^{\circ}$ with the LOS. The chance of viewing a filamentary cloud in this orientation is about 8 per cent. However, \citet{zhe21} have shown that there are a significant number of filamentary substructures identified in Orion A making an angle close to the horizontal orientation on the POS (see their figure 8). Therefore, it is possible that some filamentary substructures in OMC-4 have small inclination angles between the LOS and their axes.

\begin{figure*}
\includegraphics[angle=0,scale=0.8]{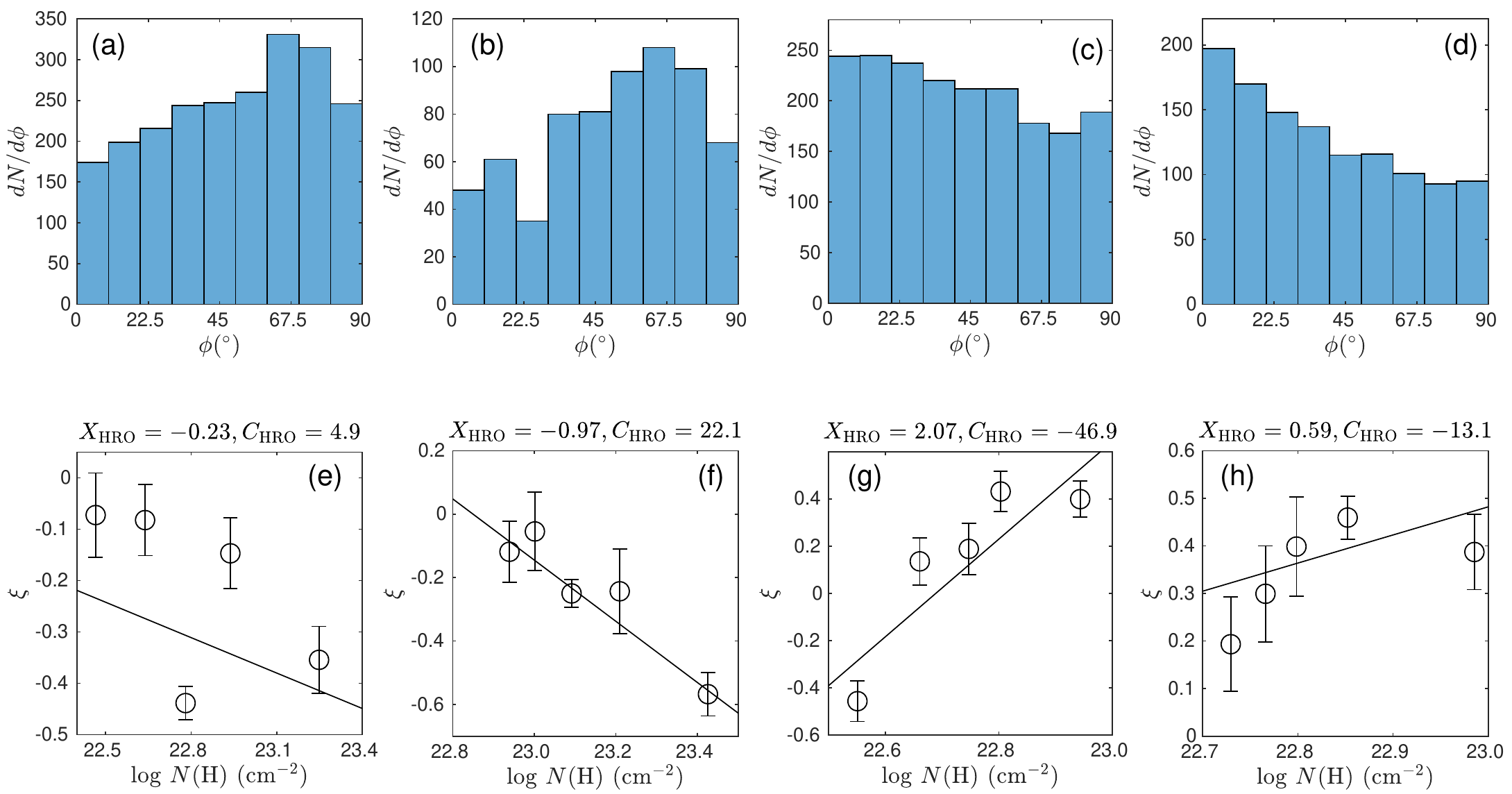}
\caption{The top rows of panels are the HROs of the angles $\phi$ between the inferred magnetic field vectors and the tangents of the column density contours for (a) the entire map of region OMC-3, (b) main cloud of region OMC-3, (c) entire map of region OMC-4, and (d) main cloud of region OMC-4. The bottom row of figure panels are the corresponding relations of shape parameter $\xi$ as the function of column density. Best $\chi^2$-fitting parameters are shown at the top of the panels.
\label{fig:Ngrad_B}}
\end{figure*}

We can see a denser elongated dense clump near the middle of the Figure \ref{fig:simcloud}b with lower density gas extending further to the upper-right and the lower-left. We assume that the cloud is optically thin and that we can see through the cloud along the LOS in the plotting. This structure is the new projection of the main filamentary cloud in Figure \ref{fig:simcloud}a. The projected magnetic field is still highly uniform but now pointing close to the long axis of the elongated dense clump.
\citet{doi20} showed that there is a probability of 14 per cent to observe a parallel B-field with the filament after assuming a random orientation between both of them in a 3D space. The picture in \ref{fig:simcloud}b resembles the polarization map of OMC-4 in Figure \ref{fig:polarization}. Since the simulated filamentary cloud has several dense clumps, the projection in Figure \ref{fig:simcloud}b appears to have more structures than the dense clump C4 in Figure \ref{fig:OMC-4_dcf}b. Therefore, the filament, where C4 is located, may not be very long and may have only one dense clump. The larger than unity mass-to-flux ratio in the observed region is mainly the result of the elongated dense clump C4. If the volume density of C4 is actually lower than expected, the mass-to-flux ratio of this region could be smaller and the entire observed region is near magnetically critical or even subcritical. In fact, the number of YSOs in this region is substantially fewer than that in the OMC-3 region (see Figure \ref{fig:polarization}). Using the data from \citet{fur16}, the total mass of the envelops around 2500 au of YSOs in the observed OMC-3 region is $12.3 \msun$, significantly larger than the $0.55 \msun$ in OMC-4, even though the total gas mass in the observed OMC-3 region is only 1.66 times of that in OMC-4 region. No YSOs are located inside the dense clumps and no prestellar cores \citep{sal15} are located in the main cloud in OMC-4. These evidences point to the conclusion that the observed OMC-4 region is stable from gravitational collapse, in contrast to the physical state in OMC-3.

\subsection{Comparison with Zeeman observations}

DCF measurements of the magnetic field are less accurate than Zeeman measurements, so comparison is worthwhile. Unfortunately, there are no Zeeman observations of OMC-3 or OMC-4, so we must compare with an average Zeeman value. Based on the Zeeman data of \citet{cru10}, \citet{li15} determined that the average field in molecular clumps is $0.19\, n_5^{0.65}$~mG. The corresponding median POS field is $0.16\,n_5^{0.65}$~mG \citep{li22}. By comparison, the average value of $B_0/n_5^{0.65}$ in the main cloud and dense clumps in OMC-3 is 0.11~mG, a little smaller than the Zeeman value. On the other hand, the average value of $B_0/n_5^{0.65}$ in the main cloud and four dense clumps of OMC-4 is 0.19~mG, quite close to the expected value. Given the small sample size, the fact that the Zeeman value is based on an average over molecular clouds in very different conditions, and the uncertainties in the DCF method, the Zeeman and DCF values of the field are in reasonably good agreement. 

\section{HRO Analysis}
\label{sec:hro}

Histogram of relative orientations (HROs) \citep{sol13,sol17} provide additional information on how dynamically important the magnetic field is in the formation of filamentary molecular clouds. Let $\phi$ be the angle between a inferred magnetic field orientation and the tangent of the column density contour so that $0^\circ \le \phi \le 90^\circ$. In Figure \ref{fig:Ngrad_B}a and \ref{fig:Ngrad_B}b, the histograms of these angles in the entire map and the main cloud of region OMC-3 are shown, respectively. There are more pixels at larger $\phi$. Figure \ref{fig:Ngrad_B}c and \ref{fig:Ngrad_B}d show the histograms of the entire map and the main cloud of the observed region in OMC-4. The distributions of $\phi$ are significantly different from that in OMC-3. There are more pixels at smaller $\phi$. As defined in \citet{sol17}, the shape parameter is
\beq
\xi = \frac{A_0-A_{90}}{A_0+A_{90}},
\label{eq:eta}
\eeq
where $A_0$ is the area under the histogram of $\phi$ values for 0$^\circ\leq\phi\leq 22.5^\circ$ and $A_{90}$ is the area for 67.5$^\circ\leq\phi\leq 90^\circ$. A negative value of $\xi$ means that the magnetic field vectors tend to be perpendicular rather than parallel to the surface contours. Physically, it is easier for gas to flow along the field than perpendicular to the field when the field is dynamically important. Furthermore, a long, slender filament can accrete gas more easily on its sides than at its ends when the field is roughly perpendicular to the long axis as observed in many filamentary clouds. This accounts for the observation that the dense regions in many molecular clouds show magnetic fields that tend to be perpendicular to contours of the surface density \citep{pla16}. In the bottom row of panels in Figure \ref{fig:Ngrad_B}, we plot the functions of $\xi$ vs $N({\rm H})$ in the corresponding cases as in Figure \ref{fig:Ngrad_B}a to \ref{fig:Ngrad_B}d, respectively. Using the same convention of symbols in \citet{pla16}, the best $\chi^2$-fitted slope, $X_{\rm HRO}$, and intercept, $C_{\rm HRO}$, of $\xi$ vs $N({\rm H})$ relation for each available column density ranges are shown at the top of the figure panels. 

\begin{figure*}
\includegraphics[angle=0,scale=0.78]{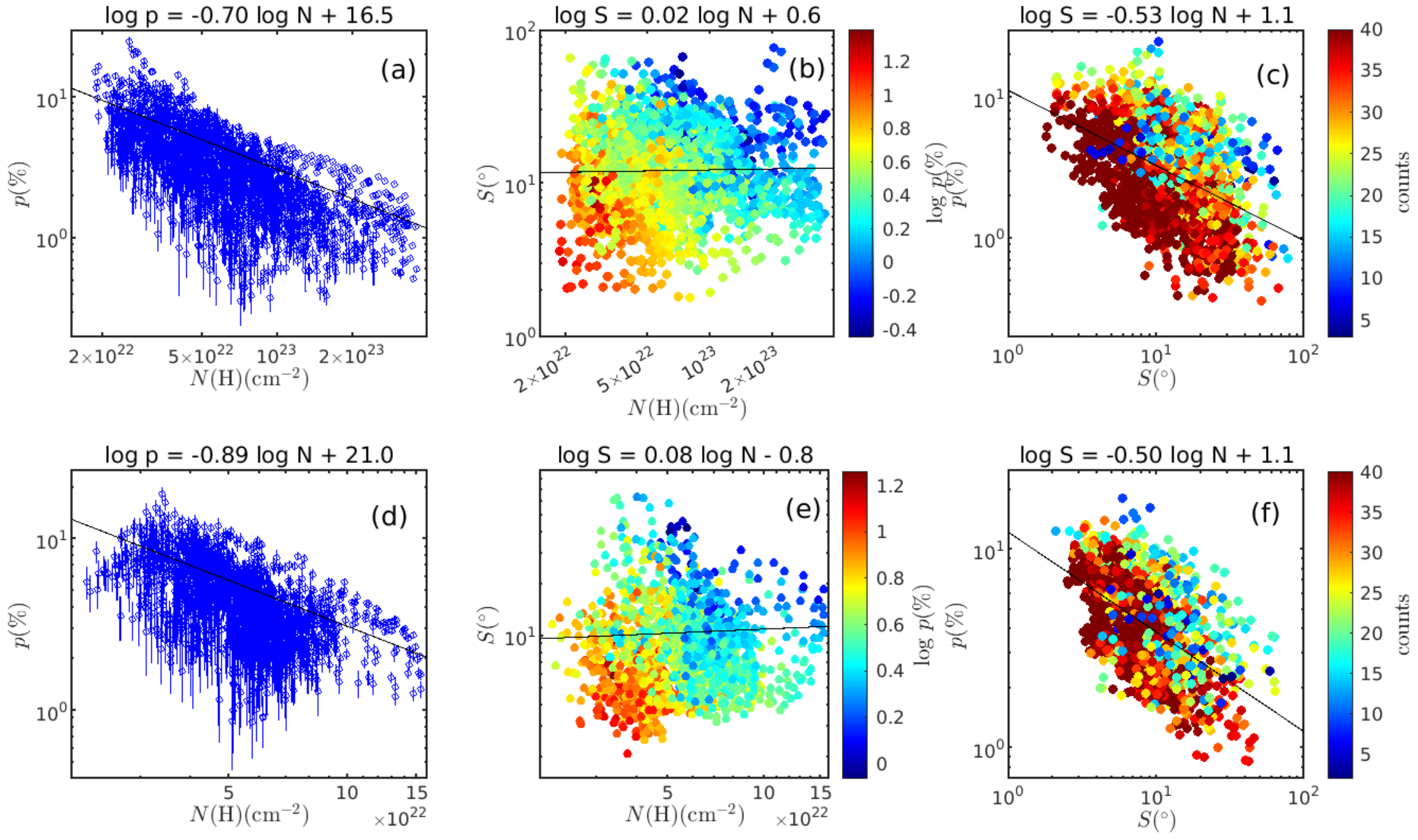}
\caption{(a) Polarization fraction $p$ versus column density $N({\rm H})$, (b) polarization-angle dispersion function $S$ versus column density $N({\rm H})$, and (c) polarization fraction $p$ versus polarization-angle dispersion function $S$ for region OMC-3. The bottom panels (d), (e), and (f) are the same as (a), (b), and (c) but for region OMC-4, respectively. The best fitted relations are marked at the top of the panels.
\label{fig:polfraction}}
\end{figure*}

For the entire map of region OMC-3, all the $\xi$ are negative. This indicates that there are more pixels having magnetic fields perpendicular than parallel to the column density contours. The magnetic field is sub-dominant. There is a trend that $\xi$ becomes more negative at higher density, indicating that the magnetic field is dynamically important during the formation of dense structures in OMC-3. The slope $X_{\rm HRO}$ is -0.23. If we treat the third data point to be an outlier and exclude the point in the fitting, $X_{\rm HRO}$ is -0.37. This trend is more clear in the main cloud of the region OMC-3 with a steeper slope of -0.97. However, for the entire map and the main cloud of the observed region in OMC-4, most of the $\xi$ are positive. Also $\xi$ increases with column density until ${\rm log} N({\rm H}) \approx 22.8$. Fluctuation in $\xi$ as a function of column density is commonly seen in many molecular clouds \citep[e.g.][]{pla16,sol17}. What is shown in Figure \ref{fig:Ngrad_B}g and h could be just the result of capturing such fluctuation. However, the positive values of $\xi$ as large as +0.46 indicate that the magnetic fields are more parallel to the column density contours. Such large $\xi$ would be seen at low column density at log $N({\rm H}) \la 21.5$ in molecular clouds \citep[e.g.][]{pla16, pat19}. The large $\xi$ at log $N({\rm H}) \la 22.8$ in the observed region in OMC-4 is quite unusual. This unusual situation may be related to the viewing angle in the observation. When the LOS is close to a filamentary cloud long axis as discussed in Section \ref{sec:projection}, we could see elongated high column density structure aligned with the uniform magnetic field. In that case, we shall see more pixels with the projected component of the magnetic field parallel to the tangents of the column density contours.

\section{Depolarization Effect}
\label{sec:depolarization}

Depolarization effect in molecular clouds, which is a trend of the decrease in polarization fraction as the column density or intensity increases, have been commonly found in polarization observations on molecular clouds \citep[e.g.][]{alv14,kwo18}. Depolarization effect has been found by \citet{mat01} and others \citep[e.g.][]{hou04,poi10} for different regions in Orion A. The depolarization effect can be described by the polarization parameter $\alpha$ in the power law relation
\beq
{\rm log}~p = \alpha ~{\rm log}~N({\rm H}) + C,
\eeq
where p is the polarization fraction, the power law index $\alpha$ is the polarization parameters, $N$(H) is the column number density, and a constant $C$. Figures \ref{fig:polfraction}a and \ref{fig:polfraction}d show the polarization fraction as a function of column density in the regions OMC-3 and OMC-4, respectively. All pixels in the high resolution HAWC+ data are included. The best fitted power law relations are shown at the top of the panels in the figure. The depolarization effect is clear. The polarization parameters for the observed OMC-3 region is -0.7 from the $\chi^2$-fitting, very close to the -0.65 obtained by \citet{mat01} for the OMC-3 clouds. \citet{poi10} find a shallower polarization parameter of -0.4 for OMC-3. By looking at Figure \ref{fig:polfraction}a and \ref{fig:polfraction}d, it seems that there could be two different populations which can be fitted with different slopes. Note that the observed OMC-3 region by HAWC+ does not cover all the dense clouds in OMC-3 as in \citet{mat01} and \citet{poi10}. Also, the polarization vectors they obtained are located at higher column density clouds, which will be similar to our case of main cloud in OMC-3 region. If we limit the $\chi^2$-fitting to only pixels in the main cloud at $N({\rm H}) \ge 0.8\times10^{23} {\rm cm}^{-2}$, the best fitted polarization parameter will be -0.62. It is not much different from the fitting using all pixels. For the observed OMC-4 region, the polarization parameter is -0.89, not too different from the OMC-3 region. There is no report on polarization parameter for OMC-4 region by others to compare with.

Depolarization effect can be a result of increasing field disorder at higher density. Local turbulence can be enhanced by gravitational collapse of dense regions. Helical structure of magnetic field around filamentary structures is also proposed to explain the depolarization effect \citep{mat01}. Other possible reasons for depolarization are suspected to be the magnetic reconnection \citep{laz99} and/or from weak radiative alignment torques \citep[RATs;][]{laz07} due to weak radiation field in dense regions. Also, dust grains can change shape at higher density due to accretion of icy mantles \citep{whi08} and results in the decrease in the degree of polarization. The tangling of field lines due to the turbulence in the region can also be a plausible cause of reduction in polarization fractions \citep{pla15, pla16}. Similar results are found in the statistical analyses of ALMA investigations, at higher spatial scales \citep{le2020}. In a recent study by \citet{hoang19}, it is suggested that RATs from the intense radiation fields can spin grain at extremely fast rotation rate. This can result into centrifugal stress which exceed the maximum tensile strength of grain material, causing the disruption of large grains into smaller fragments. \citet{hoang20} presents the detailed modeling of grain disruption towards dense cores. In the absence of the field tangling, the depolarization effect observed toward protostars are inconsistent with the RAT alignment theory but can be explained by the combined effects of grain alignment and rotational disruption by RATs. The large scattering in the relation between $p$ and $N$(H) of many different molecular clouds \citep[e.g.][]{fis16,arz21} can be the combined result of the above mechanisms operating under different conditions.

To reveal if depolarization is caused by magnetic field disorder at higher density, we can plot the relations between the polarization-angle dispersion on the beam-size (18.2 arcsec) scales and the column density, $S$ vs $N$(H), and $p$ vs $S$. Dispersion function $S$ is defined \citep{pla15} as
\beq
S\equiv\left[\Delta \psi^2({\bf x},\delta)\right]^{1/2} = \left[\frac{1}{N}\sum_{\rm i=1}^N \Delta \psi_{\rm xi}^2\right]^{1/2},
\eeq
where $\Delta \psi_{\rm xi}$ is the angle difference between the PAs of any two polarization vectors. $N$ is the number of pairs (or "counts") of polarization vectors in an annulus. In computing the dispersion function S, we sum over $\Delta \psi_{\rm xi}^2$ inside an annulus of radius $\delta = |{\boldsymbol \delta}|$ (the “lag”) and width $\delta$ around the central pixel at {\bf x}. The lag we use is the beam size of HAWC+ at Band E in the observation. In Figure \ref{fig:polfraction}b, \ref{fig:polfraction}c, \ref{fig:polfraction}e, and \ref{fig:polfraction}f, the relations between $S$ vs $N$(H) and $p$ vs $S$ are plotted for the two regions. We can see that there is basically no correlation between the dispersion function and column density in both regions. The correlation is still non-existed even when we only consider the high density pixels in the main clouds of both regions. For $p$ vs $S$, we see an anti-correlation between polarization fraction and the dispersion function, with adjusted r-squared of 0.22 and 0.29 in the fitting in OMC-3 and OMC-4 regions, respectively. This indicates that the depolarization effect is also caused by the dispersion in magnetic field along the LOS, if the dispersion is about the same on the POS. Future higher resolution polarization mapping of the two regions are required to find out if magnetic fields inside the dense clouds will have larger dispersion than the current HAWC+ data that we have.

\section{Conclusion and discussion}
\label{sec:conclusions}

We presented the results of our recent polarization observation in regions inside OMC-3 and OMC-4 using HAWC+ onboard SOFIA to investigate the physical environment and the stability of the molecular filamentary and clumpy structures in the popular Orion A region. In order to estimate the magnetic field strengths of the two regions using the classical DCF and the variant DCF/SF methods, we use the map of optical depth from \citet{lom14} obtained by using a combination of \textit{Planck} dust-emission maps, \textit{Herschel} dust-emission maps, and the 2MASS NIR extinction maps for density estimation. We examine the high-resolution $^{12}$CO(1-0), $^{13}$CO(1-0), and C$^{18}$O(1-0) molecular lines survey data of Orion A from \citet{kon18} for velocity dispersion estimation. We have performed a correlation analysis of the intensities of these three molecular lines with the polarization intensities from HAWC+ of the two regions and found that C$^{18}$O(1-0) line has the best spatial correlations with polarization intensities. In addition, the study by \citet{shi14} shows that the C$^{18}$O(1-0) line is always optically thin in the Orion A. Therefore, we use the velocity dispersion data from C$^{18}$O(1-0) for velocity dispersion estimation.

From the column density maps, the two observed regions in OMC-3 and OMC-4 are already very different. There is a well-defined massive and dense filamentary cloud in the OMC-3 but the gas structures in the observed region of OMC-4 are clumpy. The relatively lower density clumps in OMC-4 appear to have even lower density filamentary-like connections. The polarization maps obtained by HAWC+ of these two regions are also quite different.

For the OMC-3 region, we find two distinct groups of distribution of polarization angles. The smaller group is found to be mostly at the lower column density. The peaks of the two groups are separated by about $100^\circ$. It appears that there is a magnetic field associated with the dense filamentary clouds superimposed on another magnetic field associated with lower density gas. The magnetic field associated with the main filamentary cloud is mostly perpendicular $(86.8^\circ \pm 14.7^\circ)$ to the cloud long axis. The estimated field strengths on the POS of gas structures in the observed OMC-3 region are consistent with previous polarization observations. Compare with the recent observation of a smaller region in OMC-3 by \citet{zie22} using HAWC+, they deduce a little larger field strength because of using a smaller angle dispersion of polarization vectors. The observed region in OMC-3 is magnetically supercritical and strongly subvirial, indicates that gravity is dynamically dominating the region. It is consistent with the fact that there are many YSOs forming in the region. Although the region is gravitationally dominant from the virial parameters and mass-to-flux ratios at different size scales, magnetic field is dynamically important in the cloud formation process from the HROs analysis. The estimated field strengths on the POS at different size scales vary by about an order of magnitude as shown in Table \ref{tab:properties}.

From the estimated field strengths of gas structures in the observed OMC-4 region, the cloud is near magnetically critical and sub-critical locally inside the cloud, except an elongated dense clump. The inferred magnetic field is highly uniform in this region and the field appears to be parallel to the dense elongated structure, which is unexpected from most of the polarization observations of dense filamentary clouds that the field is usually roughly perpendicular to the long axis of a dense cloud structure. From our HROs analysis, we find that the shape parameters, $\xi$, of the HROs have large positive values, which is common for low density gas at column density below log $\log(N({\rm H})) = 21.5$ but not at log $\log(N({\rm H})) > 22.5$ as in the observed OMC-4 region. In comparison with the dense clumps in OMC-3, the larger velocity dispersion of dense clumps in OMC-4 indicates that the LOS depths of the dense clumps may be significantly larger than the projected width of the dense clumps by a factor of 3.2. Using our filamentary cloud simulation \citep{li19}, we propose an explanation that we may be looking closely along the long axes of some low density filamentary structures in OMC-4. In our experiment, by rotating a simulated filamentary cloud so that the angle between the LOS and cloud axis is small (in our case $23^\circ$), we can see a dense elongated structure similar to the densest elongated dense clump in the observed OMC-4 region. Using this orientation, the field lines are also parallel to the long axis of the elongated structure. In the simulation, the true 3D magnetic field is mostly perpendicular to the long axis of the filamentary cloud. If this is the case in the observed OMC-4 region, the volume densities of the clumpy structures in OMC-4 would be lower. The general nature of magnetically subcritical or near critical of this region is consistent with much fewer YSOs in the region.

We find a clear depolarization effect in both observed regions. In the case of polarization fraction versus column density, the polarization parameters are similar in OMC-3 and OMC-4, -0.70 and -0.89, respectively. The polarization parameter of the observed OMC-3 region is consistent with the result from \citet{mat01} but larger than the result from \citet{poi10}. Polarization observations at the OMC-4 region have not been done, thus our $214 \mu$m HAWC+ polarization observations represent the first ones of this kind. The anti-correlation between the polarization fraction and the polarization-angle dispersion function at the HAWC+ beam-size scales of the two regions indicates that the disorder of magnetic field contributes to the depolarization. Future higher resolution polarization mapping of the two regions will provide information on whether the magnetic fields will have large dispersions in the filamentary substructures and the dense clumps.

\section*{Acknowledgments}
Support for this research is based on observations made with the NASA/DLR Stratospheric Observatory for Infrared Astronomy (SOFIA) under the 08\_0027 Program. SOFIA is jointly operated by the Universities Space Research Association, Inc. (USRA), under NASA contract NNA17BF53C, and the Deutsches SOFIA Institut (DSI) under DLR contract 50 OK 0901 to the University of Stuttgart. We thank Chris McKee, Tie Liu, Shuo Kong, and Junhao Liu for a number of very helpful discussions, especially Shuo Kong on helping us on his CO observational data. We also thank the referee for many helpful suggestions that greatly improve the paper. Support for this research was provided by NASA through a NASA ATP grant 80NSSC20K0530 (RIK \& PSL). This work was performed under the auspices of the U.S. Department of Energy by Lawrence Livermore National Laboratory under Contract DE-AC52-07NA27344 (RIK). This document has been assigned the document release number LLNL-JRNL-832665-DRAFT. This work used computing resources from an award from the Extreme Science and Engineering Discovery Environment (XSEDE), which is supported by the National Science Foundation grant number ACI-1548562, through the grant TG-MCA00N020, computing resources provided by an award from the NASA High-End Computing (HEC) Program through the NASA Advanced Supercomputing (NAS) Division at Ames Research Center, and an award of computing resources from the National Energy Research Scientific Computing Center (NERSC), a U.S. Department of Energy Office of Science User Facility located at Lawrence Berkeley National Laboratory, operated under Contract No. DE-AC02-05CH11231.

\section*{Data Availability}
The processed HAWC+ data in FITS format is available at CDS via anonymous ftp to cdsarc.u-strasbg.fr (130.79.128.5) or via https://cdsarc.unistra.fr/viz-bin/cat/J/MNRAS.

\label{lastpage}

\end{document}